\documentclass[12pt,bibliography]{article} 
 
\usepackage[]{amsmath}
\usepackage[]{graphicx}
\usepackage[]{latexsym}
\usepackage{color}
\usepackage{geometry}
\usepackage{subfigure}
\usepackage{lscape}

\usepackage{hyperref}
\usepackage[latin1]{inputenc}                                
\usepackage{color}                                           
\usepackage{array}                                           
\usepackage{longtable}                                       
\usepackage{calc}                                            
\usepackage{multirow}                                        
\usepackage{hhline}                                          
\usepackage{ifthen}      
\usepackage{pdfpages}

\def\beq{\begin{equation}} 
\def\eeq{\end{equation}} 
\def\barr{\begin{array}} 
\def\earr{\end{array}} 
\def\beqa{\begin{eqnarray*}} 
\def\eeqa{\end{eqnarray*}} 
 
\font\mybb=msbm10 at 12pt 
\def\bb#1{\hbox{\mybb#1}}

\def\b1 {\bb{1}} 
\def\ket#1{| #1 \rangle} 
\def\bra#1{\langle #1 |} 
 
\def\lesssim{\mathrel{\hbox{\rlap{\hbox{\lower4pt\hbox{$\sim$}}}\hbox{$<$}}}} 
\def\gtrsim{\mathrel{\hbox{\rlap{\hbox{\lower4pt\hbox{$\sim$}}}\hbox{$>$}}}}

\makeatletter
\def\equalsfill{$\m@th\mathord=\mkern-7mu
  \cleaders\hbox{$\!\mathord=\!$}\hfill
  \mkern-7mu\mathord=$}
\makeatother

\def\hksqrt{\mathpalette\DHLhksqrt}
\def\DHLhksqrt#1#2{\setbox0=\hbox{$#1\sqrt{#2\,}$}\dimen0=\ht0
  \advance\dimen0-0.2\ht0
  \setbox2=\hbox{\vrule height\ht0 depth -\dimen0}%
{\box0\lower0.4pt\box2}}

\DeclareMathOperator{\erfc}{erfc}
  
\makeatletter
\def\blfootnote{\xdef\@thefnmark{}\@footnotetext}
\makeatother

\begin{document} 

\vspace*{-.6in} \thispagestyle{empty}
\begin{flushright}
\end{flushright}
\vspace{.2in} {\Large
\begin{center}
{\bf Vector meson production at low $x$ from \\ gauge/gravity duality}
\end{center}}
\vspace{.2in}
\begin{center}
Miguel S. Costa$^1$\blfootnote{miguelc@fc.up.pt}, Marko Djuri\'c$^1$\blfootnote{djuric@fc.up.pt}, Nick Evans$^2$\blfootnote{evans@soton.ac.uk}
\\
\vspace{.3in} 
\emph{
$^1$ Centro de F\'\i sica do Porto,
Departamento de F\'\i sica e Astronomia,
Faculdade de Ci\^encias da Universidade do Porto, 
Rua do Campo Alegre 687,
4169--007 Porto, Portugal
\ \\ 
$^2$ STAG Research Centre, Physics and Astronomy, University of Southampton, Highfield, Southampton, SO17,1BJ, UK}

\end{center}

\vspace{.3in}

\begin{abstract}
We use gauge/gravity duality to study vector meson ($J/\Psi, \rho_0, \Omega, \Phi$) production  in electron-proton scattering, in the limit of high center of mass  energy
at fixed momentum transfer, corresponding to the limit of low Bjorken $x$, where the process is dominated by pomeron exchange. 
Our approach considers the pomeron at strong coupling, described by the graviton Regge trajectory in AdS space with a hard-wall to mimic confinement effects.
Both the proton and vector mesons are described by simple  holographic wave functions in AdS.
This model agrees with HERA H1 data with a $\chi^2$ per degree of freedom below one on  total cross-sections, and below two on  differential cross-sections,  
confirming the success of previous studies that model low $x$ DIS and DVCS using gauge/gravity duality. 

\end{abstract}

\newpage

\setcounter{page}{1}


\vfill\eject

\section{Introduction}

\begin{figure}[t!]
\begin{center}
\includegraphics[scale=0.5]{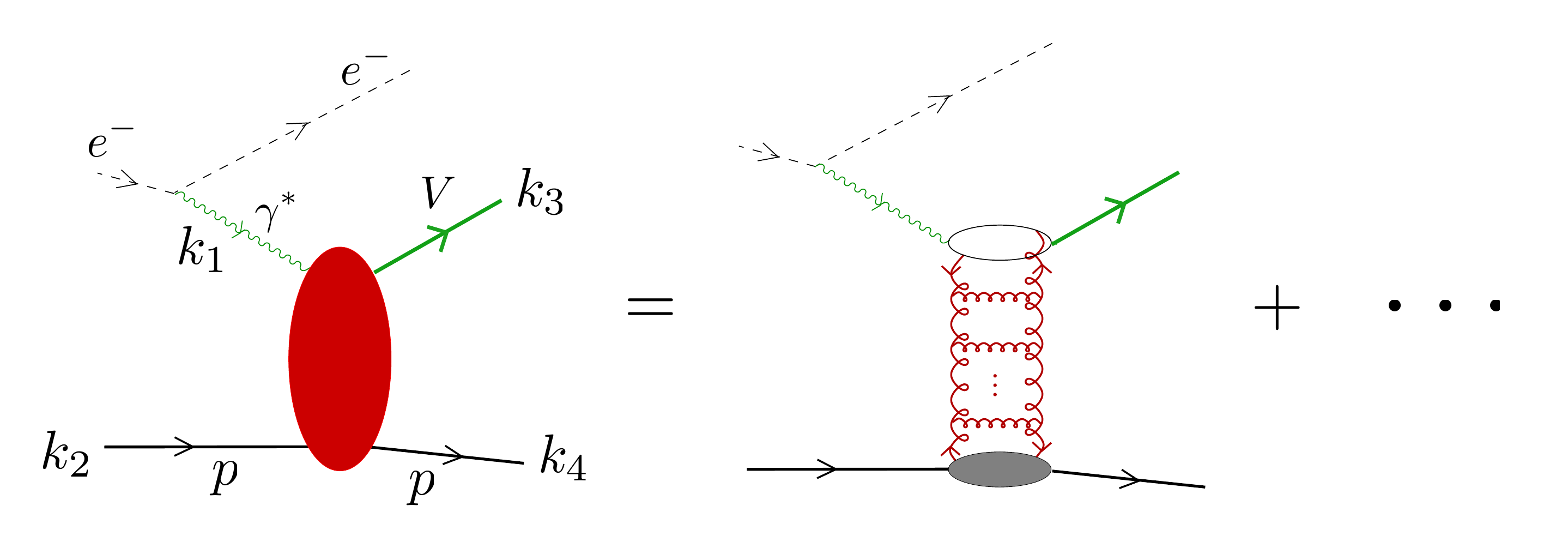}
\caption{{\small The process of vector meson production analysed in this paper. An off-shell photon of momenta $k_1$ (with $k_1^2=Q^2$) interacts with a proton $p$ of momenta $k_2$, leading to a final state of a scattered proton and a vector meson  $V$. At low $x$ this process is dominated by the exchange of the pomeron, which at weak coupling can be described 
by the BFKL hard pomeron.}}
\label{VMPdiagram}
\end{center}
\end{figure}

Vector meson production (VMP) is one of the diffractive processes studied in electron-proton collisions at HERA. It is conceptually similar to deeply virtual Compton scattering (DVCS), 
but instead of an outgoing photon a vector meson is produced. The vector mesons have the same $J^{PC}$ values as the photon 
(i.e. $1^{--}$), so the process is kinematically similar. The key difference comes from the vector mesons' structure functions. 
Here we will study processes where the final state is a $\rho, \phi, J/\psi$ or $\Omega$. Figure \ref{VMPdiagram} shows the process considered in this paper.

In the limit of high center of mass  energy at fixed momentum transfer, corresponding to the limit of low Bjorken $x$, VMP is dominated by the exchange 
of the pomeron Regge trajectory between the photon and the proton. In particular, at large virtuality $Q^2$ of the incoming photon, 
diagrams of order $[\alpha_s \ln (1/x)]^n$ can be resummed in perturbation theory and the process is described by the exchange of the hard pomeron 
\cite{Fadin:1975cb,Kuraev:1977fs,Balitsky:1978ic}, 
as shown schematically in Figure \ref{VMPdiagram}.
A number of authors
 \cite{Martin:2007sb,Dosch:2006kz,Marquet:2007qa,Donnachie:2008kd,Fazio:2011ex,Berger:2012wx,Rezaeian:2012ji}
 have studied vector meson production using the weak coupling analysis providing a decent fit to the data. More recently, the analysis in \cite{Forshaw:2012im} uses AdS wave functions within a dipole approximation to fit $\rho$ production.

\begin{figure}
\begin{center}
\includegraphics[scale=0.42]{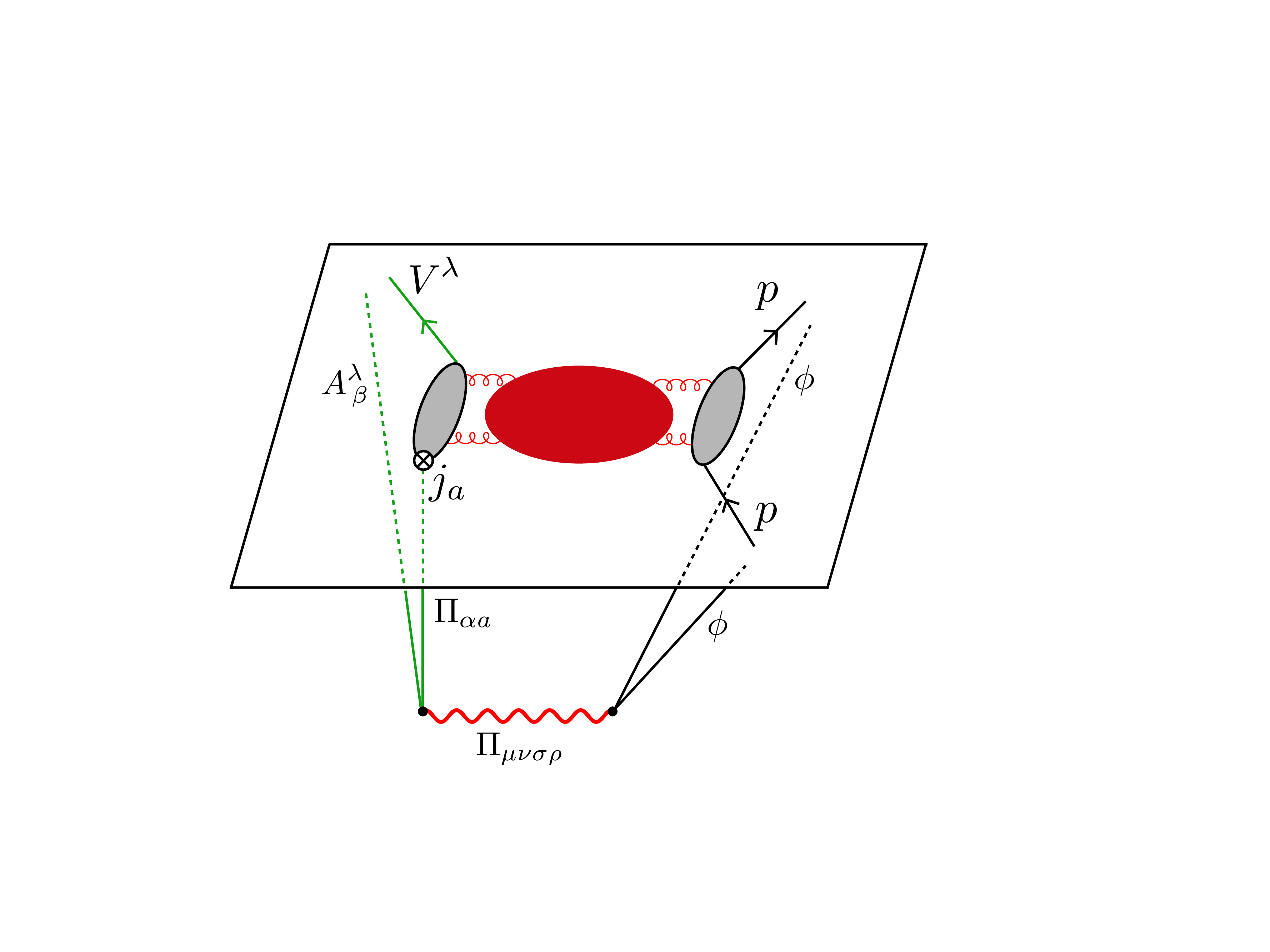}
\caption{{\small Holographic representation of the hadronic tensor needed to compute the cross section for 
VMP. 
Single pomeron exchange at strong coupling is described by the exchange of the AdS graviton Regge trajectory.
In the light-cone coordinates used in this paper the exchange is dominated by the $\Pi_{++--}$ component of the Reggeon
propagator.}}
\label{fig:feynman_vmp}
\end{center}
\end{figure}

The Gauge/Gravity duality establishes a correspondence between the pomeron Regge trajectory and the graviton Regge trajectory of the dual string theory \cite{BPST}.
This correspondence has been used to study low-$x$ QCD processes dominated by pomeron exchange such as DIS  
\cite{Hatta:2007he,BallonBayona:2007qr,BallonBayona:2007rs,Bayona:2011xj,satDIS,Albacete:2008ze,Levin:2009vj,Hatta:2009ra,Kovchegov:2009yj,Avsar:2009xf,ourBlackDisk,
Levin:2010gc,Brower:2010wf,Betemps:2010ij}  and DVCS \cite{Nishio:2011xa,Nishio:2011xz,Costa:2012fw}.
In particular, it provides a description for the strong coupling expansion of  pomeron exchange, whose intercept varies from $j_0=1$ at weak coupling to $j_0=2$ at
strong coupling. 
Experimental evidence shows that, as one increases the coupling (by varying $Q^2$ from large to small), 
the effective spin of the exchanged pomeron grows from  $j_0\sim1.1$ to $j_0\sim 1.4$ (see for instance the figure presented in the conclusion of \cite{Brower:2010wf}).
It is therefore conceivable that one may fit the data starting from the strong coupling regime, instead of the more conventional weak coupling BFKL approach.
This is done here by considering   the graviton Regge trajectory in AdS space. 
Confinement can be simply and successfully modelled by the inclusion of an infra-red hard-wall in this AdS space. 
A detailed discussion on the validity and assumptions behind this model was presented in \cite{Costa:2012fw}, here we will be mainly concerned in
computing the cross section for VMP at low $x$ using the dual AdS tree level diagram with the exchange of the graviton Regge trajectory,
as shown in Figure \ref{fig:feynman_vmp}. We then compare to HERA data.

A new key aspect of our gauge/gravity duality description of VMP, in comparison with DIS and DVCS, will be to use a very simple holographic model for the vector mesons which gives the holographic wave function of the mesons as a function of their mass. 
These wave functions are normalisable modes of the AdS $U(1)$ gauge field dual to the electromagnetic current operator
$j^a_f=\bar{\psi}_f \gamma^a \psi_f $. 
Holographic models of vector mesons include using the D3/D7 system \cite{Karch:2002sh,Grana:2001xn,Bertolini:2001qa,Kruczenski:2003be,Erdmenger:2007cm} and the Sakai-Sugimoto model \cite{Sakai:2004cn,Sakai:2005yt} - in each case a bulk gauge field is associated with the vector meson states and eigenvalues of its radial wave equation are computed with the AdS radius terminating at the scale of the constituent quark's mass. AdS/QCD is a phenomenological crystallization of these ideas \cite{Erlich:2005qh,DaRold:2005zs}.
In coordinates where the AdS metric is given by
$ds^2=(R^2/z^2)(dx^2+dz^2)$,
for each meson we will simply solve for the radial wave function of an AdS vector field  cut off at some radius $z_f$, which is determined by the physical  mass of the meson.
For most of the analysed data the overlap between the bulk-to-boundary propagator, which creates a source of the  operator $j^a_f$ that couples to 
the virtual photon, the meson wave functions and the pomeron bulk-to-bulk propagator is in the
UV region. In this region both the bulk-to-boundary propagator and the meson wave functions are determined by the canonical dimension of $j^a_f$. When the overlap moves to the QCD scale we rely on the simplest phenomenological model with a cut-off  $z_f$ for the  field dual to $j^a_f$, and also with an AdS hard-wall at a cut-off $z_0$ for the remaining fields.
For the proton we assume that its wave function is  highly peaked at a scale $z_*$. 

The methodology we use here for pomeron exchange has already been developed and tested against experimental data in previous publications 
\cite{satDIS,Brower:2010wf,Costa:2012fw}. Holographic descriptions of vector mesons are discussed in detail in \cite{Erdmenger:2007cm}. Hence, to keep this note short, we will quickly summarize the main results and new features needed for VMP, and then proceed to present the results. The reader is directed to the aforementioned sources for more information on using the AdS/CFT correspondence in the study of diffractive scattering. 

It is however important to enumerate clearly the free parameters in our model. The meson wave functions are determined in terms of the scale $z_f$, but we fix this by the observed meson mass.  Our wave function for the proton state includes the scale $z_*$ which we keep as a fit parameter. The Regge-graviton propagator 
is constructed from the conformal limit at strong coupling, 
where it depends on one parameter -- the intercept $j_0$. There is one additional parameter that enters, $g_0^2$,
which is determined by  the coupling of the pomeron to the external states (and is therefore different for each vector final state). 
A fit to the data assuming the conformal propagator hence depends on three parameters and we find already a good fit.  
A fourth parameter can be introduced to represent confinement in the propagator, a hard-wall cut off at large $z_0 > z_f$. 
We also make fits including this adjustment and find a better overall $\chi^2$ fit. 

In the results section of the paper we present numerical fits of our model to HERA H1 data for $\rho, \phi$ and  $J/\psi$ production as well as to the very limited ZEUS data for the $\Omega$. Our results are summarized in Table 1 of Section \ref{sec:Results}. The overall fit to the total cross-sections has a $\chi^2$/degree of freedom below one in each case, which confirms  the approach captures the physics of the processes. The fits to the differential cross-sections have a $\chi^2$/degree of freedom between one and two, which whilst not as good a fit as previously found for the DIS and DVCS using this approach, is still decent.

\section{Cross section for  vector meson production}
To compute the total cross section for vector meson production we need first to compute the following hadronic  tensor
\begin{equation}
W_{a}^{\ \lambda}(k_j) =  i \int d^4y\,e^{i k_1 \cdot y}\bra{k_3,\lambda; k_4 }  j_a(y)   \ket{k_2}\,,
 \label{DefHadronicTensor}
\end{equation}
where $\lambda$ is the polarization of the outgoing vector meson and we refer to Figure \ref{VMPdiagram} for the kinematics. 
Contracting with the polarization of the incoming photon,  the amplitude
for the transition between a photon of polarization $\lambda$ and a vector meson of polarization $\lambda'$ is 
\begin{equation}
W^{\lambda \lambda'}(k_j) =     (n^\lambda)^a  W_{a}^{\ \lambda'}(k_j)  \,.
\label{DefHadronicTensor2}
\end{equation}
We will average over the incoming polarizations and sum over the final ones. 
The differential cross-section is then given by
\begin{equation}
\frac{d\sigma}{dt} (x,Q^2,t)= \frac{1}{16\pi s^2}\,\frac{1}{3} \sum_{\lambda,\lambda'=1}^3 \big| W^{\lambda\lambda'} \big|^2    \,.
\label{differential}
\end{equation}
We will compute $W^{\lambda\lambda'}$ from the AdS Witten diagram sketched in Figure \ref{fig:feynman_vmp}. 
We will now discuss the key elements of such diagram.

\subsection{External kinematics}

Let us describe the kinematics of the external particles. We use 
light-cone coordinates $(+,-,\perp)$, with metric given by $ds^2= -dx^+ dx^- + dx_\perp^2$, 
where $x_\perp \in \bb{R}^2$ is a vector in  impact parameter space. For the incoming particles we take
\begin{equation}
k_1 = \left(\sqrt{s} ,- \frac{Q^2}{\sqrt{s}},0\right)\,,  
\ \ \ \ \ \ \ \ \ \
k_2 = \left(\frac{M^2}{\sqrt{s}},\sqrt{s},0\right)\,,
\label{k1k2}
\end{equation}
where $M$ is the mass of the target and the incoming off-shell photon is space-like with $k_1^2=Q^2>0$.
For the outgoing particles
\begin{equation}
k_3 =  - \left(\sqrt{s} ,\frac{q_\perp^2 + m^2}{\sqrt{s}}, q_\perp\right)\,,   
\ \ \ \ \ \ \ \ \ \
k_4 = -  \left(\frac{M^2+q_\perp^2}{\sqrt{s}},\sqrt{s},-q_\perp\right)\,,
\label{k3k4}
\end{equation}
where $m$ is the mass of the vector meson that is created.
We consider the Regge limit of large $s$ and fixed $t=- q_\perp^2$.

We also need to define the photon and vector meson polarization vectors.
Let $n_\lambda$ and $n'_{\lambda}$ be, respectively, the photon and vector meson polarization vectors.
We can use the gauge freedom to impose the conditions 
\begin{equation}
n \cdot k_1 =0\,,\ \ \ \ \ \ \ \ n' \cdot k_3 =0\,.
\end{equation}
We normalize the photon polarization such that $n^2=1$ for transverse polarizations and 
$n^2=-1$  for longitudinal polarization (the photon is space-like). For the vector meson
we have always $n'^2=1$.

In the above light-cone coordinates, the polarization of the transverse incoming photon is 
\begin{equation} \label{photonin}
n_\lambda = (0,0, \epsilon_\lambda)\,,\ \ \ \ \ \ \ \ \ \ \ \ \ \ \ (\lambda=1,2)
\end{equation}
where $\epsilon_\lambda$ is an orthogonal basis of unit vectors on $\bb{R}^2$. The incoming longitudinal photon has polarization
\begin{equation} \label{photonin2}
n_3 =   \frac{1}{Q}\left(\sqrt{s} , \frac{Q^2}{\sqrt{s}},0\right)\,,
\end{equation}
where we define $Q=\sqrt{Q^2}$. Note that  all the $n_i$ are orthogonal (and unit normalized).

For the  outgoing meson  we introduce the two (transverse) polarizations
\begin{equation}
n'_\lambda = \left( 0, 2\,\frac{\epsilon'_\lambda\cdot q_\perp}{\sqrt{s}} , \epsilon'_\lambda\right)\,,\ \ \ \ \ \ \ \ \ \ \ \ \ \ \ (\lambda=1,2)
\label{VectorTransverse}
\end{equation}
where  $\epsilon'_\lambda$ is an orthogonal  basis of unit vectors on $\bb{R}^2$.
This polarization is transverse, in the sense that in the Regge limit the leading component is on the transverse space $\bb{R}^2$.
For the other polarization we take
\begin{equation}
n'_3 =\frac{1}{m} \left(  \sqrt{s}, \frac{-m^2 + q_\perp^2}{\sqrt{s}} ,q_\perp \right)\,.
\label{VectorLongitudinal}
\end{equation}
This polarization is longitudinal, in the sense that in the Regge limit the leading component is light-like and parallel to the meson momenta.
Again all   $n'_i$ are orthogonal (and unit normalized).

\subsection{External state AdS wave functions}
\label{sec:WaveFunctions}

In the gravity dual description bound states are associated with eigenmode wave functions in the holographic $z$ direction. 
The vector mesons for example are described by a normalizable mode of the AdS $U(1)$ gauge field dual to the electromagnetic current operator
$j^a_f=\bar{\psi}_f \gamma^a \psi_f $.  The non-normalizable mode is dual to this operator's  source (which couples to the virtual photon when we
include the QED coupling to the quarks).
To fix the asymptotic normalization of the mode we must impose a large $z$ (IR) boundary condition on the solution. We simply include a ``fermion hard-wall" at the scale $z_f \sim m_f^{-1}$ for each fermion flavour and impose Neumann boundary conditions on the field at the wall. The value of $z_f$ is phenomenologically fixed by the measured vector meson mass and the model contains no free parameters in the mesonic sector. 

We write the normalizable mode for the AdS gauge field describing the vector mesons as
\begin{equation}
A^{\lambda}_{\mu}(X) = \int \frac{d^4k}{(2\pi)^4}  \, n'^{\lambda}_\mu \, e^{ik\cdot x} A(z,k)\,,
\label{MesonWaveFunctionAnsatz}
\end{equation}
where $X=(z,x)$ are the usual Poincar\'e coordinates in AdS. In the gauge $D_\mu A^\mu=0$, the field equation $D^2 A_\mu=0$
is solved by
\begin{equation}
A(z,k)= \frac{\sqrt{2}}{\xi J_1({\xi})} \,mz J_1(mz)\,, \hspace{1cm} k^2=-m^2\,,
\label{MesonWaveFunction}
\end{equation}
with the polarisation vector  satisfying
\begin{equation}
n'_z=0\,,\ \ \ \ \ \ \ \ n'_a k^a =0\,,
\end{equation}
where the boundary polarisation $n'_a$ is given by either (\ref{VectorTransverse}) or (\ref{VectorLongitudinal}).
We have taken the normalizable Bessel function solution, so that the wave function 
falls off asymptotically as $z^{2}$, as required for a massless vector field dual to the vector operator $j^a_f $ of dimension 3.
The constant  $\xi=2.4048...$ in the expression (\ref{MesonWaveFunction}) for $A(z,k)$  is the first zero of the Bessel funcion $J_0$
and arises from imposing Neumann boundary conditions on the field at the 
cut-off scale $z_f$ defined by the quark cut off scale $z_f \sim 1/m_{q_f}$.
The relation between the meson mass and the corresponding quark cut off is 
then $m = \xi /z_f$.
The overall constant in $A(z,k)$ follows from the normalization condition
\begin{equation}
\int_0^{z_{f}} \frac{dz}{z} | A(z,k)|^2  = 1\,.
\end{equation}

To compute the hadronic tensor (\ref{DefHadronicTensor}) we need to compute the expectation value of the 
operator $j^a_f$ in the state defined by the incoming proton, and outgoing proton and vector meson. Thus 
we need to include the bulk-to-boundary propagator of the same gauge field, given by its non-normalizable mode,
which is  dual to a source of  $j^a_f$ (see Figure \ref{fig:feynman_vmp}) and behaves  asymptotically as $z^0$. 
Denoting the propagator between the bulk point $X=(z,x)$ and the boundary point $y$ by
$\Pi_{\mu a}(X,y)$, in the
gauge $D_\mu A^\mu=0$
we have
\begin{equation}
\Pi_{za}(X,y) =0\,,\ \ \ \ \ \ \
\Pi_{ba}(X,y) = \int \frac{d^4k}{(2\pi)^4}  \, e^{ik\cdot (x-y)} \Pi_{ba} (z,k)\,,
\end{equation}
with
\begin{equation}
\Pi_{ba} (z,k) = \sqrt{C\, \frac{\pi^2}{6}}   \left(  \eta_{ba} -\frac{k_bk_a}{k^2} \right) \,Q z K_1(Qz)\,, \hspace{1cm} k^2=-Q^2\,.
\label{BbProp}
\end{equation}
Note that this is the solution in a conformal theory that extends to infinite $z$. Using the same simple model as above, we should also introduce 
Neumann boundary conditions at the IR wall located at $z=z_f$, as described in \cite{Hong:2004sa}. 
However, a simple analysis of the overlap between the vector meson wave function and the bulk-to-boundary propagator 
shows that this modification is only important when the vector meson mass $m$ is of the same order as the off-shellness 
$Q$. This does not happen for most of the data points here considered. 
Indeed we have found changes in the IR boundary conditions of the bulk-to-boundary propagator to have little impact on the fits.

The constant $C$ in (\ref{BbProp}) is fixed by the normalization of the current operator two-point function given by
$\langle j_a(y) j_b(0)\rangle = (C/y^6) (\eta_{ab} - 2 y_ay_b/y^2)$. In free theory, for a quark of charge
$q_f$, we have  $C=3 \alpha q_f^2/\pi^3$, where $\alpha=e^2/(4\pi)$ is the fine structure constant. 
To fix a reference value for this normalization constant we assume the flavour content of our vector mesons is given by $\rho = {1 \over \sqrt{2}} (u\bar{u} - d \bar{d})$,  
$\Omega = {1 \over \sqrt{2}} (u\bar{u} + d \bar{d})$, 
$\Phi = s \bar{s}$ and  $J/\psi = c \bar{c}$. We then have  $C_\rho = C_\Omega = 5 \alpha/ (6 \pi^3)$,  $C_\Phi = \alpha/(3 \pi^3)$ and $C_{J/\psi} = 4 \alpha/ (3 \pi^3)$.  
Since over the kinematical range 
of VMP here considered the theory is not free, our fits will also allow for changes from this reference values.

At the opposite vertex of the t-channel exchange  a proton scatters and in   
principle one should put the AdS wave functions for both incoming and outgoing protons.
For example, we could describe the proton by the normalizable mode of a scalar in AdS with mass squared of 9/4. 
The solution, $\phi(z) \propto z^{3/2}\left(-\sin{z}-3\cos{z}/z + 3 \sin{z}/z\right)$, would fall off as $z^{9/2}$ matching the dimension of $\psi_f^3$.  This behaviour would extend up to some scale $z_*$ associated with the dynamical mass of the proton's constituents.
In fact, since the  proton wave function is sufficiently fast falling at small $z$, we 
will take simply a delta function localized at a scale
$z_*$. More concretely we consider
\begin{equation}
\Phi(z) = |\phi(z)|^2 = z^3 \delta( z - z_*)\,,
\label{target}
\end{equation}
which satisfies the normalization condition
\begin{equation}
\int \frac{dz}{z^3}\,  |\phi(z)|^2 = 1\,.
\end{equation}
We will leave $z_*$ as a free parameter in our fits.
An additional benefit of this approach is that the $z$ integration at the vertex can be done analytically rather than numerically, greatly reducing the numerical computation time for the process. This is a considerable saving across many fits. In fact we did try some sample runs using the full proton wave function but found only a very small change in the fit parameters.

\subsection{Witten diagram for graviton exchange}

In the Regge limit, the amplitude for $\gamma^* p \rightarrow V p$, computed from the dual 
Witten diagram in AdS, is dominated by t-channel exchange of the  graviton Regge trajectory. 
Let us first consider the  limit of very large 't Hooft coupling, where graviton exchanges dominates
the other massive strings in this trajectory. 
In this case, for the above external Regge kinematics, standard Feynman rules in AdS give
\cite{ourSW,ourEikonal,Brower:2007qh,Bartels:2009sc,ourDIS}
\begin{align}
 W_{a}^{\ \lambda}(k_i)   = & \  \kappa^2 
\int \,dl_\perp\, e^{i q_\perp \cdot l_\perp }
\int   \frac{dz}{z^3}\,\frac{d\bar{z}}{\bar{z}^3} \,
2 \partial_{1\left[-\right.}\Pi_{\left.\alpha\right] a}(z,k_1)\,g^{\alpha\beta}(z)\, 2\partial_{3\left[-\right.}A^{\lambda}_{\left.\beta\right]}(z,k_3)
\nonumber
\\&
ik_{2+}\phi(\bar{z},k_2)\, ik_{4+}\phi(\bar{z},k_4)   16   z\bar{z}  \,\Pi_\perp(L)
\,,
\end{align}
where we used the notation
\begin{equation}
2 \partial_{1\left[-\right.}\Pi_{\left.\alpha\right] a}(z,k_1) = \sqrt{C\, \frac{\pi^2}{6}} \times
\left\{
\begin{array}{ll}
2i 
k_{1\left[-\right.}\eta_{\left. c \right] a} \,Qz K_1(Qz)\,,& \alpha=c
\\ 
\left( \eta_{-a} -\frac{k_{1-}k_{1a}}{Q^2}\right)Q^2z K_0(Qz)\,,  \ \ \  & \alpha=z
\end{array}
\right.,
\end{equation}
and
\begin{equation}
2\partial_{3\left[-\right.}A^{\lambda}_{\left.\beta\right]}(z,k_3) = \frac{\sqrt{2}}{\xi J_1({\xi})}  \times
\left\{
\begin{array}{ll}
2i
k_{3\left[-\right.}n'^{\lambda}_{\left. d \right] }
\,mz J_1(mz)\,, \ \ \  & \beta=d
\\ 
- n'^\lambda_{-}   \,m^2z J_0(mz)\,, & \beta=z
\end{array}
\right..
\end{equation}
We have discussed in the previous section the form of the 
bulk-to-boundary propagator of the AdS vector field and also of the normalizable modes
dual to the vector meson and to the proton.

The above amplitude already takes into account that
for the Regge kinematics here considered 
the component $\Pi_{++--}$ of the graviton propagator 
dominates the exchange, and can be integrated along the light cone directions to give a
scalar propagator $\Pi_\perp(L)$ on the AdS transverse space $H_3$ of mass squared 3 \cite{ourSW}, with
\begin{equation}
\cosh L=\frac{z^2+\bar{z}^2+l_\perp^2}{2z\bar{z}}\,.
\end{equation}

Next we contract the amplitude $W_{a}^{\ \lambda'}(k_i) $ with the incoming photon polarization $n^\lambda$, and then consider
all possible polarizations $\lambda$ and $\lambda'$. It turns out that  the
 non-zero contributions preserve helicity, i.e. $W_{LT}=0=W_{TL}$. A simple computation shows that
\begin{align}
W_{TT} &= (n^\lambda)^a\,  W_{a}^{\ \lambda'}(k_j) =\left( \epsilon_\lambda\cdot \epsilon_{\lambda'} \right) Qm \,W_1\,,\ \ \ \ \ \ (\lambda,\lambda'=1,2)
\\
W_{LL}&= (n^\lambda)^a \, W_{a}^{\ \lambda'}(k_j) = - Qm \,W_0\,,\ \ \ \ \ \  \ \  \ \ \ \,  \ \ \ (\lambda=\lambda'=3)
\end{align}
where
\begin{equation}
W_n  =  2 i s  \int dl_\perp\,
 e^{iq_\perp\cdot l_\perp} 
 \int \frac{dz}{z^3}\, \frac{d\bar{z}}{\bar{z}^3} \,
\Psi_n(z)\, \Phi (\bar{z})  \,
\left[  i \,\frac{\kappa^2}{2}\,S\,\Pi_\perp(L) \right] \,,
 \label{Wn}
\end{equation}
with $S=z\bar{z}s$,
 \beq
 \Psi_n(z) = - \left( \sqrt{\frac{C\pi^2}{6}} \,z^2 K_n(Qz) \right) \left(  \frac{\sqrt{2}}{\xi J_1({\xi})} \,z^2 J_n(mz)\right)\,,
 \eeq
 and $\Phi(\bar{z}) $ is given in (\ref{target}) above.
  
\subsection{Exchange of graviton-Regge trajectory}

So far we have evaluated the Witten diagram with a single graviton exchange which is valid for very large 't Hooft coupling $\lambda$.
To extend this result to smaller coupling/string tension, including the correction from string states in the leading Regge trajectory,
we first need to realise that the amplitude (\ref{Wn}) can be written in terms of a conformal amplitude 
${\cal B}(S,L)$. Taking care of the external polarizations, one can then show that the amplitude $W_n$ defined in (\ref{Wn})
becomes
\begin{equation}
W_n  =  2 i s  \int dl_\perp\,
 e^{iq_\perp\cdot l_\perp} 
 \int \frac{dz}{z^3}\, \frac{d\bar{z}}{\bar{z}^3} \,
\Psi_n(z)\, \Phi (\bar{z})  \,
{\cal B}(S,L) \,.
 \label{WnGeneral}
 \end{equation}
 This general form relies only on conformal invariance, as shown in \cite{ourDIS} and reviewed at length in \cite{Costa:2012fw}.
 At very large 't Hooft coupling, introducing the AdS phase shift $\chi(S,L)=i{\cal B}(S,L)$, the Witten diagram for single graviton 
 exchange in the Regge limit has
\begin{equation}
\chi(S,L) =   - \frac{\kappa^2}{2} S \,\Pi_\perp(L)\,.
\end{equation}

In this paper we are interested in the limit of large  't Hooft coupling $\lambda\gg 1$, but with sufficiently high energies such that  
$\sqrt{\lambda}/ \ln S \ll 1$. In this limit all fields in the graviton Regge trajectory contribute to the amplitude \cite{Brower:2007xg}, and we have 
\begin{equation}
{\cal B}(S,L) =  \,g_0^2  \left( 1+i \cot\Big(\frac{\pi \rho}{2}\Big) \right)  
\,(\alpha' S)^{1-\rho}\,  \frac{e^{-\frac{L^2}{\rho\ln (\alpha' S)}}}{  (\rho \ln (\alpha' S))^{3/2}} \,\frac{ L }{ \sinh L} \,,
\label{AmplitudeFinal}
\end{equation}
where 
\begin{equation}
\alpha' S=\frac{z\bar{z}s}{\sqrt{\lambda}}\,,\ \ \ \ \ \ \ \ \ \ 
\rho= 2-j_0=\frac{2}{\sqrt{\lambda}}\,.
\label{alpha'S}
\end{equation}
In this equation the 't Hooft coupling is defined from the AdS radius as $\lambda = R^4/\alpha'^2$, where $\alpha'$
is the tension of the dual QCD string\footnote{In ${\cal N}=4$ SYM, $\lambda=g^2_{YM}N$ is also the 't Hooft coupling defined
from perturbation theory. On the other hand, in the kinematic range analysed in this paper, where we start by 
considering that in this range QCD is approximately conformal, just as in the weak coupling BFKL analysis, 
the precise relation between $\lambda = R^4/\alpha'^2$ and the 't Hooft coupling defined from perturbation theory is not known.
This would mean finding the dual QCD string.}.
The coupling $g_0^2$ is related to the impact factors of the external states. 
The resulting graviton-Regge amplitude depends 
on two parameters: the intercept $j_0$ (or equivalently the above defined  't Hooft coupling $\lambda$) and the coupling $g_0^2$. 
Note that, as an overall coefficient of the Feynman diagram, $g_0^2$ also soaks up any freedom in the 
normalization constant $C$ of the $U(1)$ gauge field bulk-to-boundary propagator. In other words, the fits to data will 
fix only the combination $\sqrt{C}g_0^2$, and for this reason we  shall fix, when presenting the results, $C$ to the reference values given in Section 
 \ref{sec:WaveFunctions}.

As so far presented the graviton-Regge, or the BPST pomeron \cite{BPST}, trajectory exchanged in AdS is that of a conformal theory. 
In QCD we expect that confinement will enter in the IR and cut off any deep IR contribution to the process. The simplest model of that effect is just to include a hard-wall in $z$ at $z_0$ on the Pomeron propagator. We will introduce this extra parameter into a second fit to the data below also. In this model, the propagator for the Pomeron is modified, due to the different boundary conditions in the differential equation that defines it \cite{BPST}. In our analysis, following \cite{Brower:2010wf}, we use the approximation (with $\tau = \log(\alpha' S)$)
\begin{equation}
  \label{eq:chi_hw}
  \chi_{hw}(\tau,l_\perp,z,\bar{z}) = C(\tau,z,\bar{z})\,D(\tau ,l_\perp)\, \chi^{(0)}_{hw}(\tau,l_\perp,z,\bar{z})\,,
\end{equation}
where   
\begin{equation}
  \label{eq:D}
  D(\tau,l_\perp) = \min\left(1,\frac{ \exp[-  m_1 l_\perp  -  (m_0-m_1)^2 l_\perp^2 / 4 \rho \tau ]}{\exp[-  m_1 z_0  - ( m_0-m_1)^2  z_0^2 / 4 \rho \tau ]}    \right) ,
\end{equation}
is an exponential cutoff at large $l_\perp$, known to be present asymptotically and determined by the first glueball masses $m_0$ and $m_1$, and
\begin{equation}
  \chi^{(0)}_{hw}(\tau,l_\perp,z,\bar{z}) =  \chi_{c}(\tau ,l_\perp,z,\bar{z}) + {\cal F}(\tau ,z,\bar{z}) \,\chi_{c}(\tau,l_\perp,z,z_0^2/\bar{z})\,.
\end{equation}
 In the above $\chi_{c}(\tau,l_\perp,z,\bar{z})$ is derived from the conformal kernel, equation \eqref{AmplitudeFinal}. This approximation stems from the fact that the hard-wall kernel can be shown to have the above form at $t=q_\perp^2=0$, i.e. with $l_\perp$ integrated over while setting $q_\perp = 0$, and then extrapolating to $l_\perp$. $C(\tau,z,z')$, is a normalization function, independent of $l_\perp$, which ensures the $t = 0$ result is retained. For details see \cite{Brower:2010wf,Costa:2012fw}. The function  
\begin{equation}
  \label{eq:F}
  \mathcal{F}(\tau,z,\bar{z}) = 1-4\hksqrt{\pi\tau}\,e^{\eta^2}\erfc(\eta)\,, \ \ \ \ \ \ \  \eta = \frac{-\log(z\bar{z}/z_0^2)+4\tau}{\hksqrt{4\tau}}\,,
\end{equation}
is set by the boundary conditions at the wall and represents the relative importance of the two terms and therefore confinement. This function varies between $-1$ and $1$, approaching $-1$ at either large $z$, which roughly corresponds to small $Q^2$, or at large $\tau$ corresponding to small $x$. It is therefore in these regions that confinement is important.

\begin{figure}[t!]
\begin{center}
\includegraphics[scale=0.75]{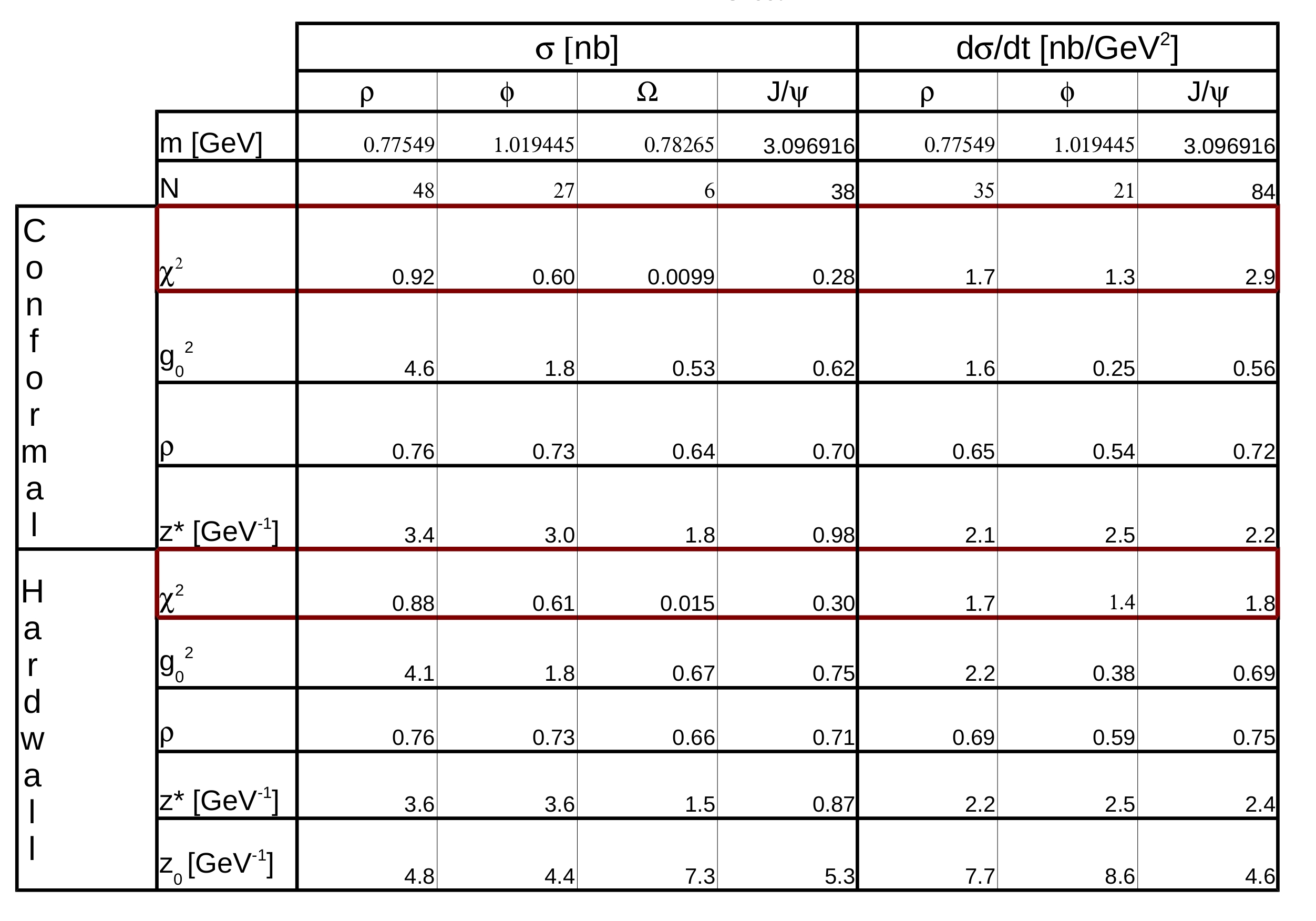}
\end{center}
\vspace{-0.3cm}
Table 1: {\small Output data for our fits showing the number of experimental points, $N$, the $\chi^2$ per degree of freedom for the fit and the best fit parameter values for 
$j_0 = 2-\rho$ (the intercept), 
$g_0^2$ (a vertex factor fitted for each meson state, with respect to reference values of the normalisation constants $C$ given in 
Section  \ref{sec:WaveFunctions}), 
$z^*$ (the IR scale characterizing the proton wave function) and, for the hard-wall model  fits,  $z_0$ (an IR confinement cut-off).}
\end{figure}

\section{Results}
\label{sec:Results}

\begin{figure}[t!]
\begin{center}
\includegraphics[scale=0.425]{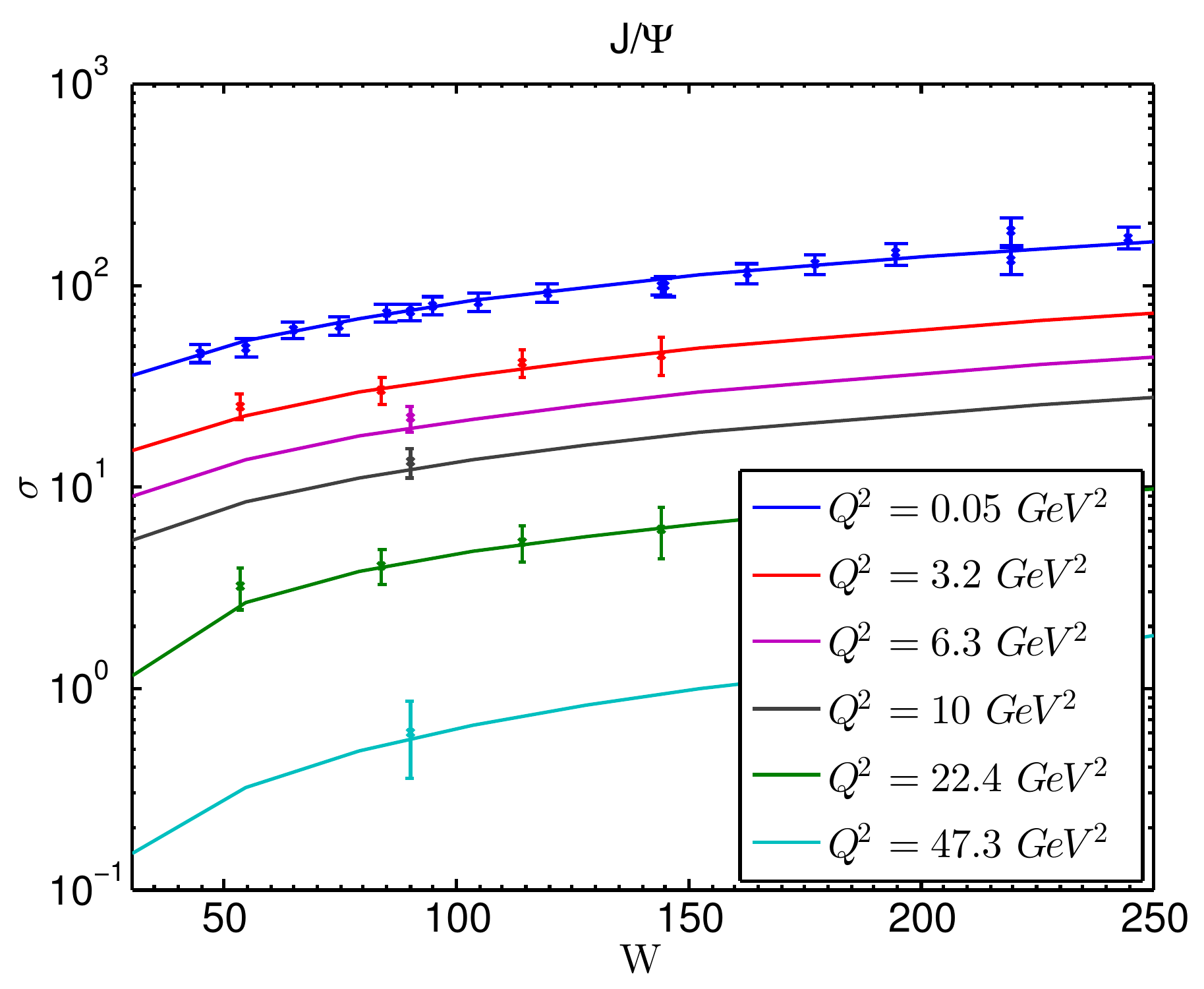}
\includegraphics[scale=0.425]{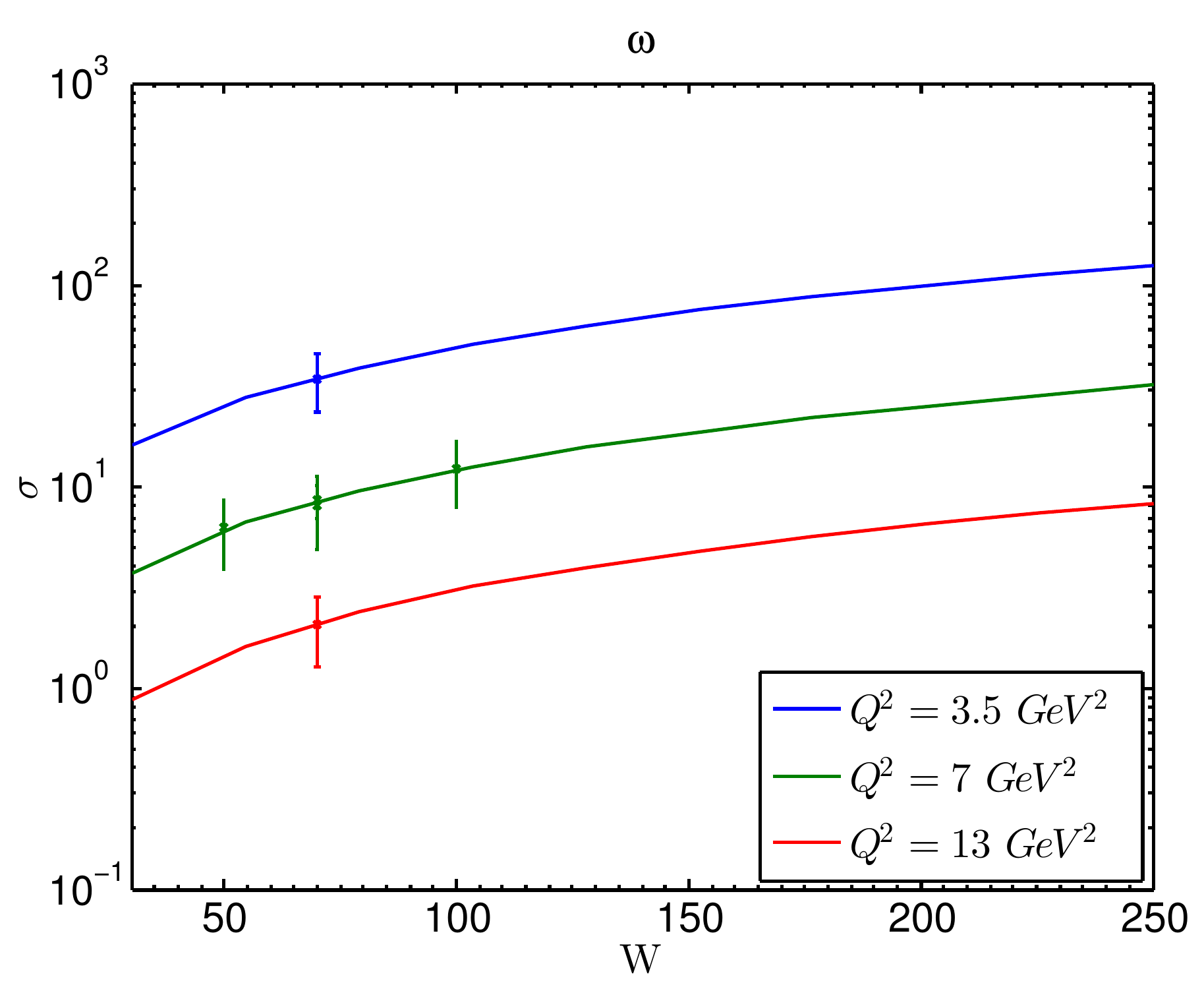}\\
\includegraphics[scale=0.425]{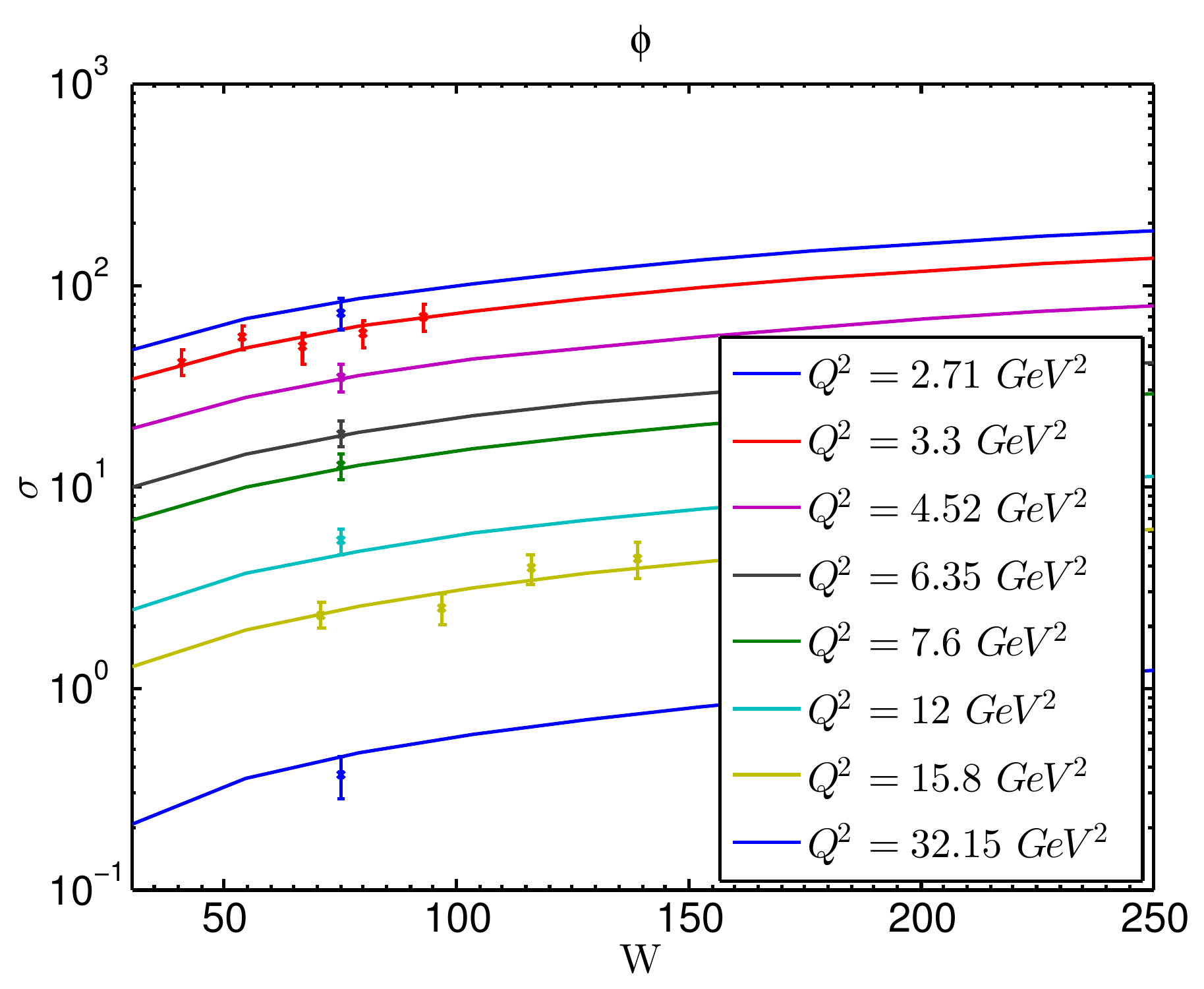}
\includegraphics[scale=0.425]{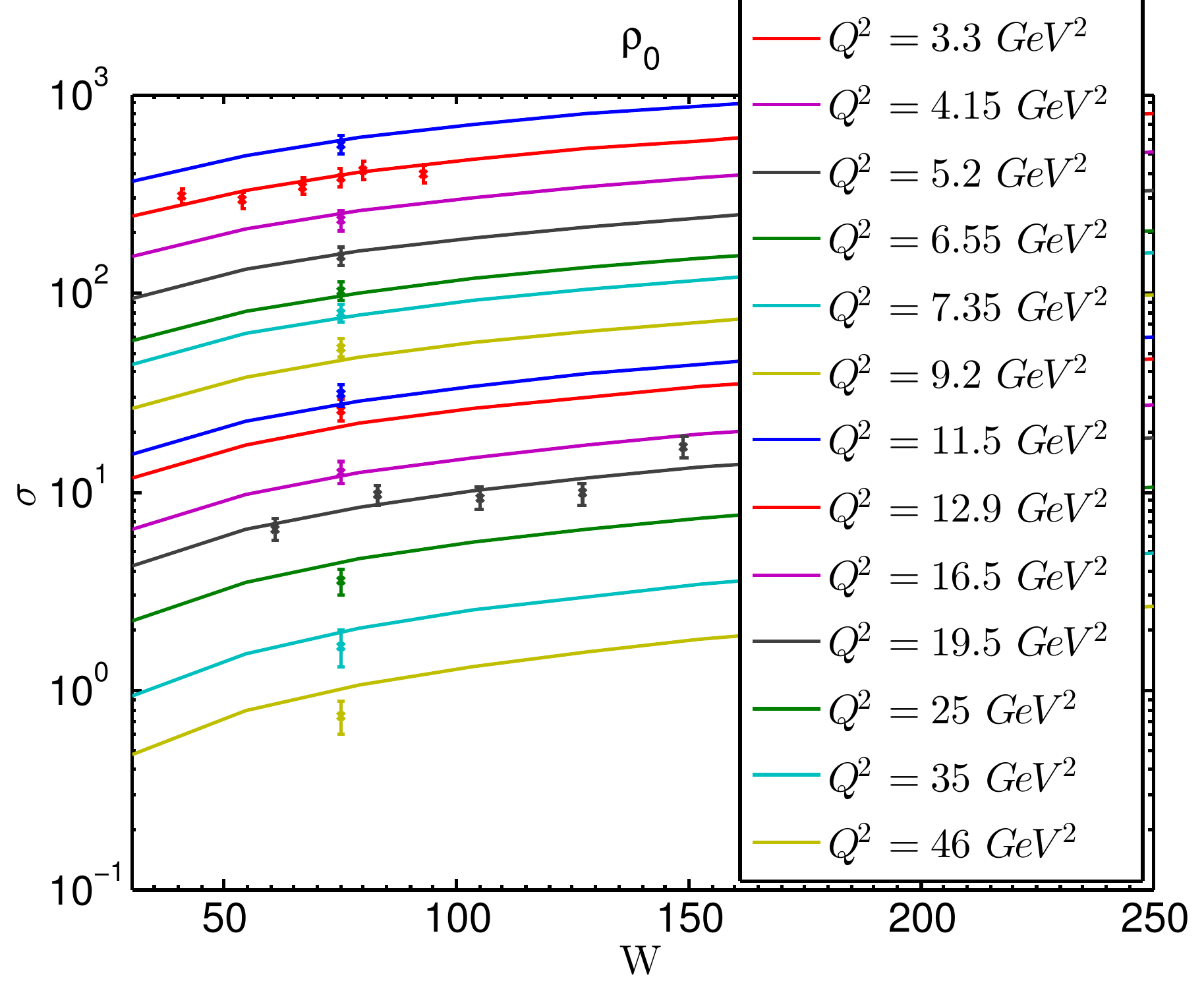}
\caption{{\small Cross sections using the conformal model.}}
\label{fig:sig_conf}
\end{center}
\end{figure}

Our tree level AdS Feynman diagram provides a full description of the low $x$ vector meson production in terms of the parameters 
$j_0 = 2-\rho$ (the intercept), $g_0^2$ (a vertex factor fitted for each meson state), $z^*$ (the
IR scale characterizing the proton wave function) and if we choose $z_0$ (an IR confinement cut off). We will now perform a fit to the data collected at HERA by the H1 collaboration \cite{Aktas:2005xu,Aaron:2009xp}. 

All of the data is at small $x$ ($<0.01$). Note that the $\rho$, $\phi$ and $\Omega$ have a mass close to each other, ranging from $\sim 0.78 - 1.02 \ {\rm GeV}$, whereas the $J/\Psi$ has a significantly higher mass at $3.09 \ {\rm GeV}$. This means that it is precisely for the $J/\Psi$ that the flavour cut-off scale becomes 
closer to the probe scale $Q$, therefore  indicating that details of the AdS model for this vector meson will be 
be more important. 


In Table 1 we see a summary of all our fits. We show fits to the full cross-section and the differential cross-sections for each process. $N$ labels the number of available data points. We list the $\chi^2$ per degree of freedom in the fit and the best fit values of the parameters.
Although data is available from the ZEUS collaboration as well \cite{Chekanov:2004mw,Breitweg:2000mu,Chekanov:2005cqa,Chekanov:2007zr}, we find that in fitting the differential cross section we obtain better fits using the H1 data. This is most pronounced in the fit for the $J/\Psi$ meson where fitting the H1 differential cross section we get the results from Table 1, while fitting just the ZEUS data we get a $\chi^2_{d.o.f.} = 9.57$. Fitting both H1 and ZEUS together for this meson (but without taking into account the normalization differences between the data sets) we get a $\chi^2_{d.o.f.} = 6.74$. For the remaining mesons fitting the data from both collaborations together gives only a slightly worse $\chi^2_{d.o.f.}$\footnote{For individual data sets, H1 gives a better fit in each case.},  but to be consistent throughout we chose to present the fits and plots for just the H1 data\footnote{The only exception is the cross section for the $\Omega$ meson, where we could only find ZEUS data.}. If the measurements of both collaborations are properly statistically combined in the manner of \cite{Aaron:2009aa}, with corresponding  fits given in \cite{Brower:2010wf} for the case of DIS, it would be possible to revisit the fits here presented. It is also interesting to note that the H1 data set for $J/\Psi$ goes to the lowest value of $Q^2$, hence giving us the most direct probe of our hard-wall model near the location of the cutoff.

\begin{figure}[t!]
\begin{center}
\includegraphics[scale=0.425]{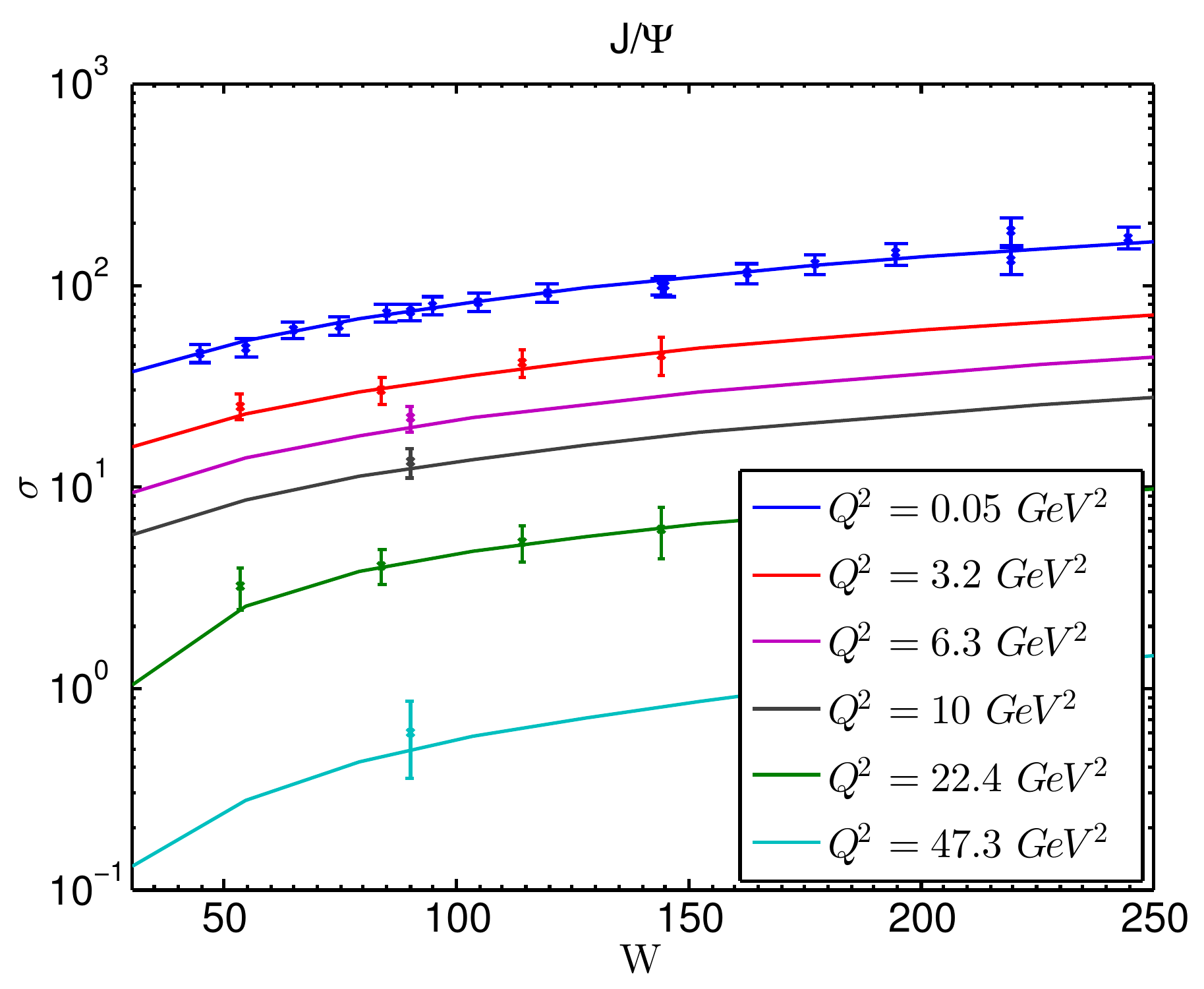}
\includegraphics[scale=0.425]{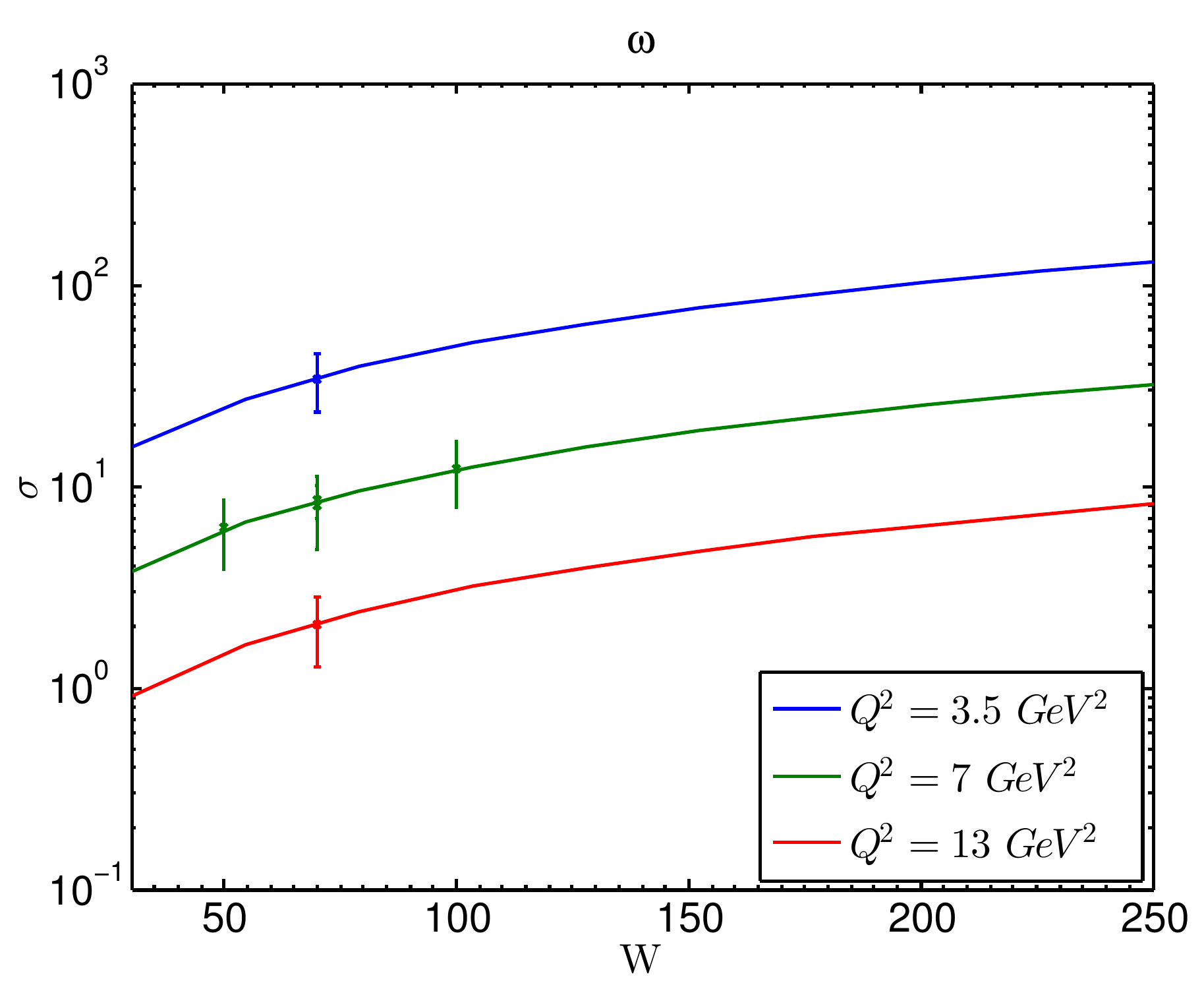}\\
\includegraphics[scale=0.425]{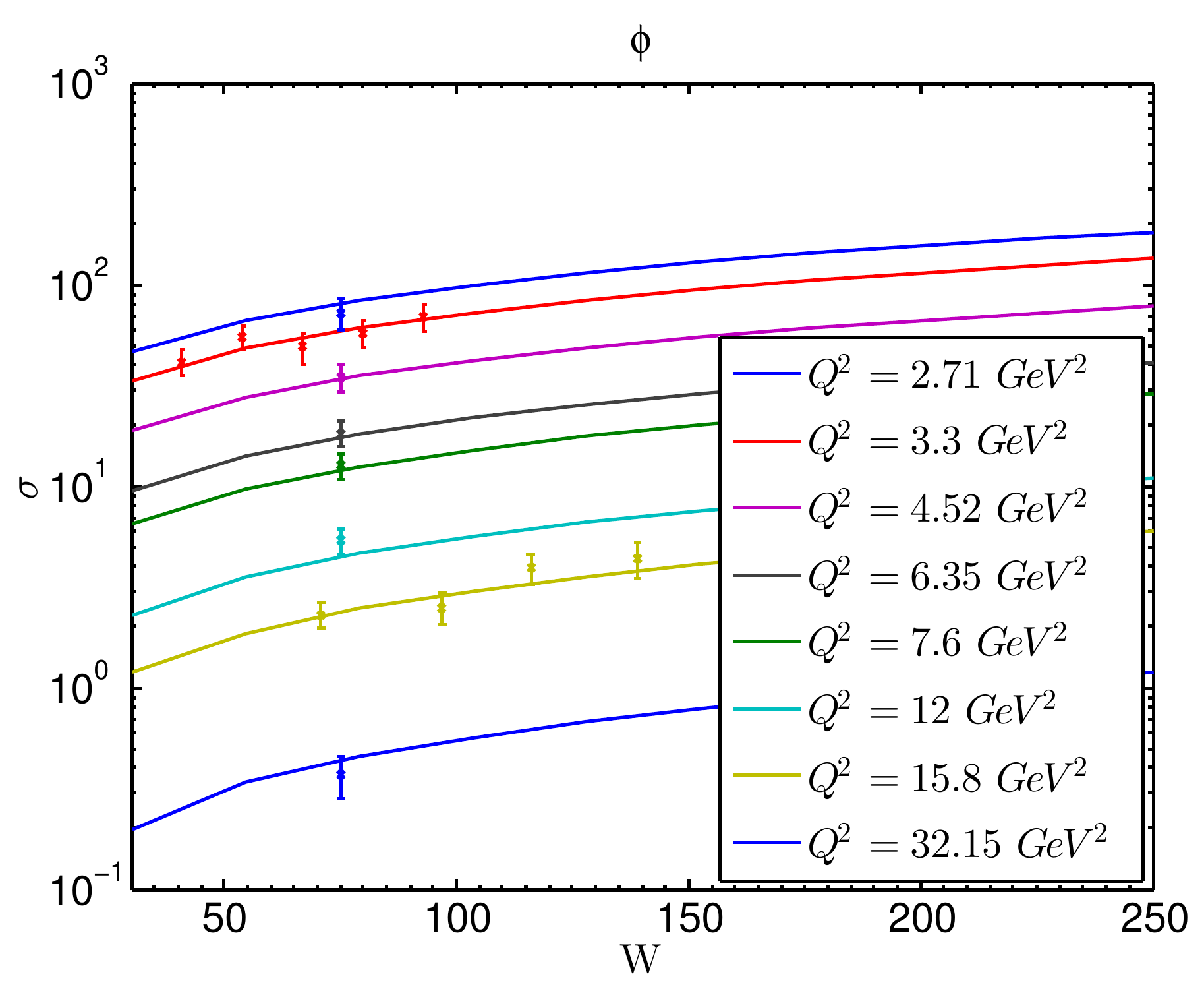}
\includegraphics[scale=0.425]{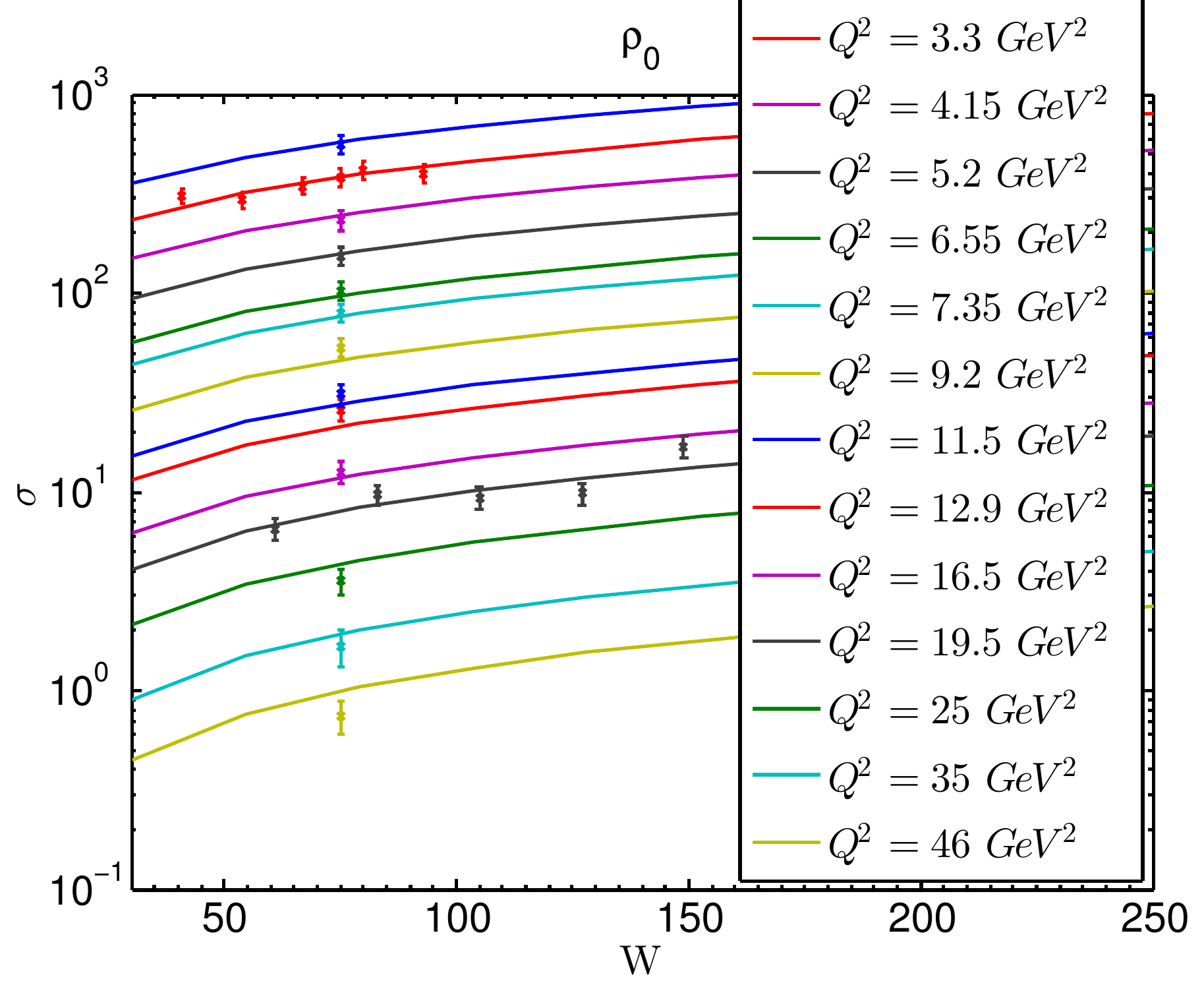}
\caption{{\small Cross sections using the hard-wall model.}}
\label{fig:sig_hw}
\end{center}
\end{figure}


\begin{figure}[t!]
\begin{center}
\includegraphics[scale=0.425]{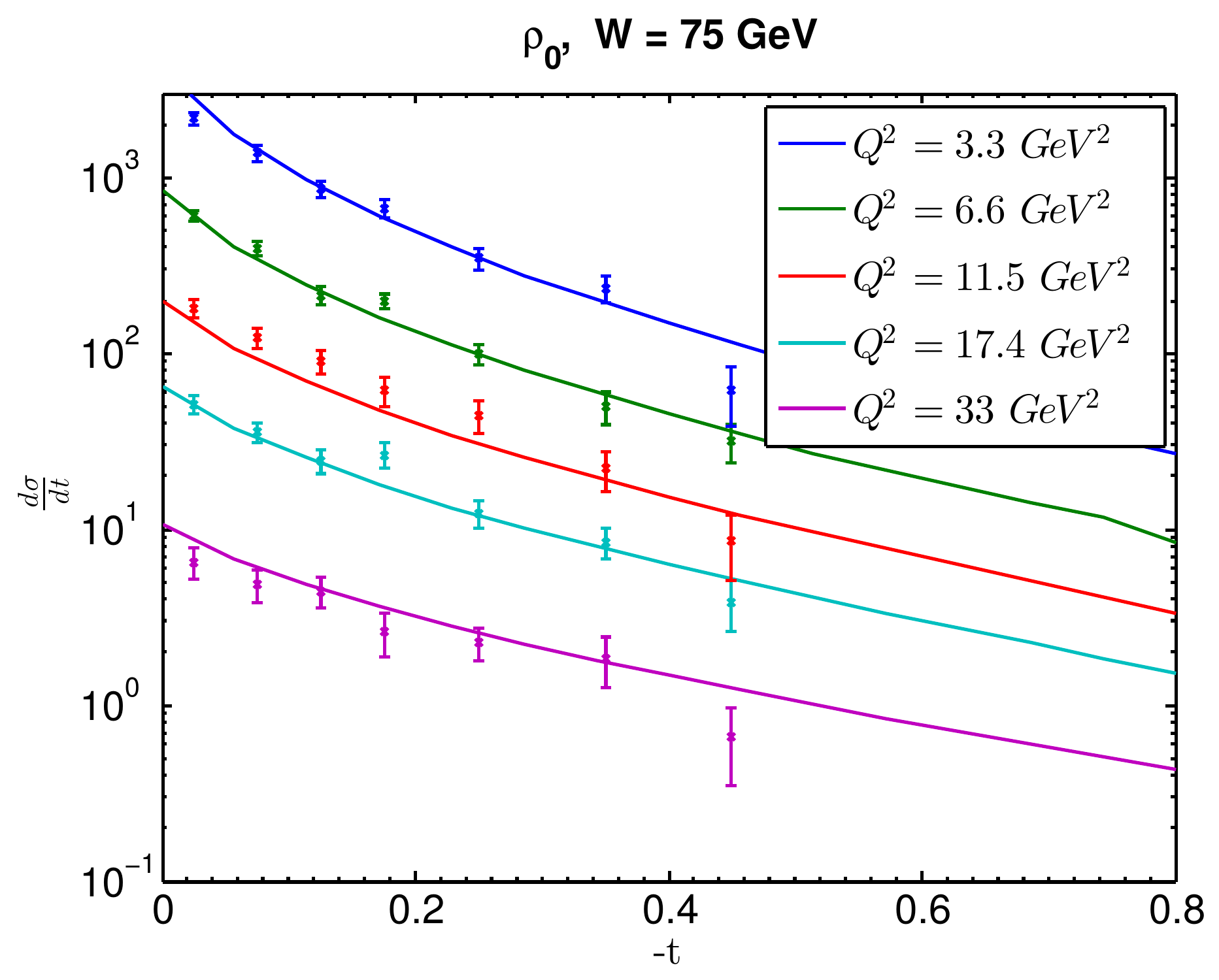}
\includegraphics[scale=0.425]{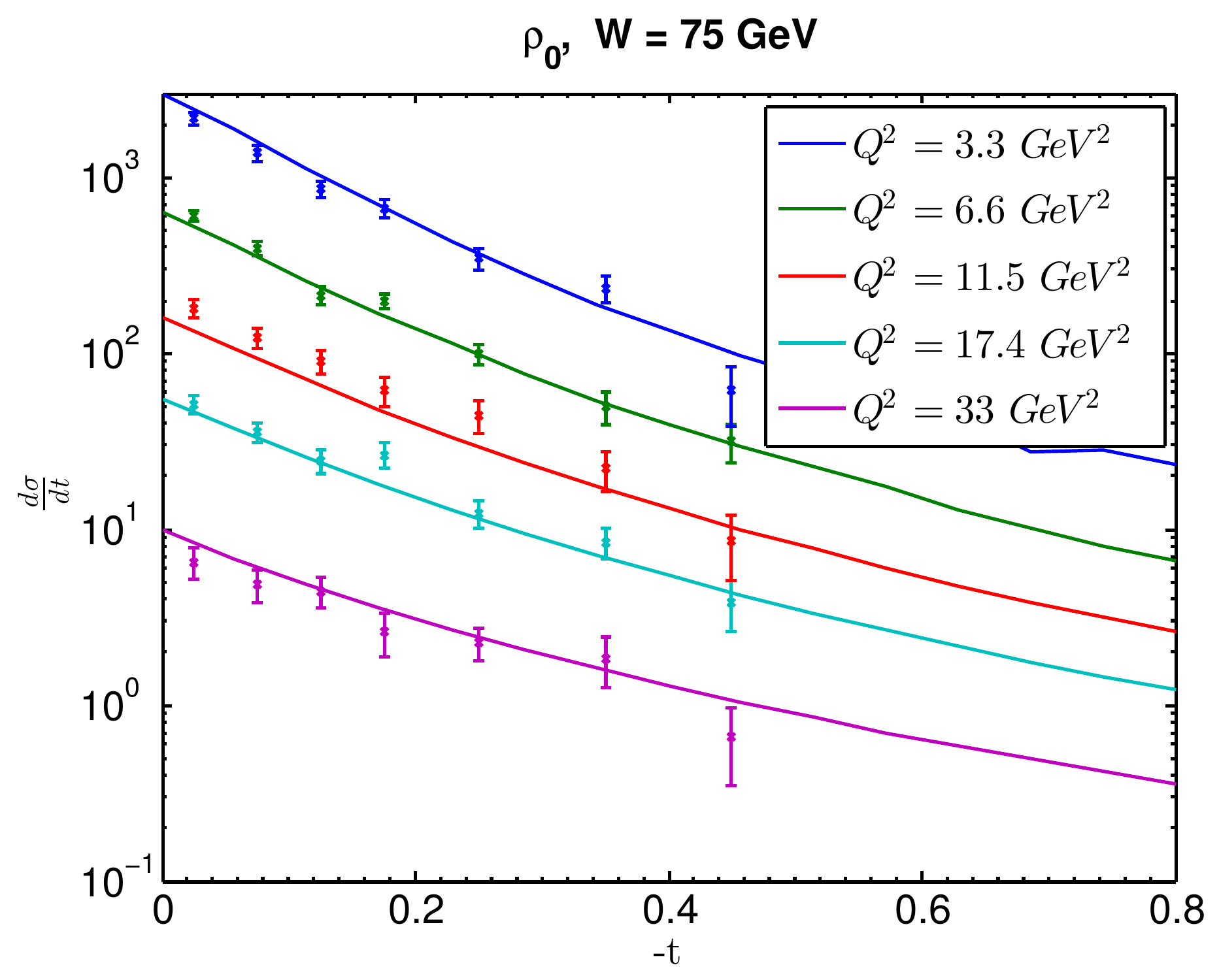} \\
\includegraphics[scale=0.425]{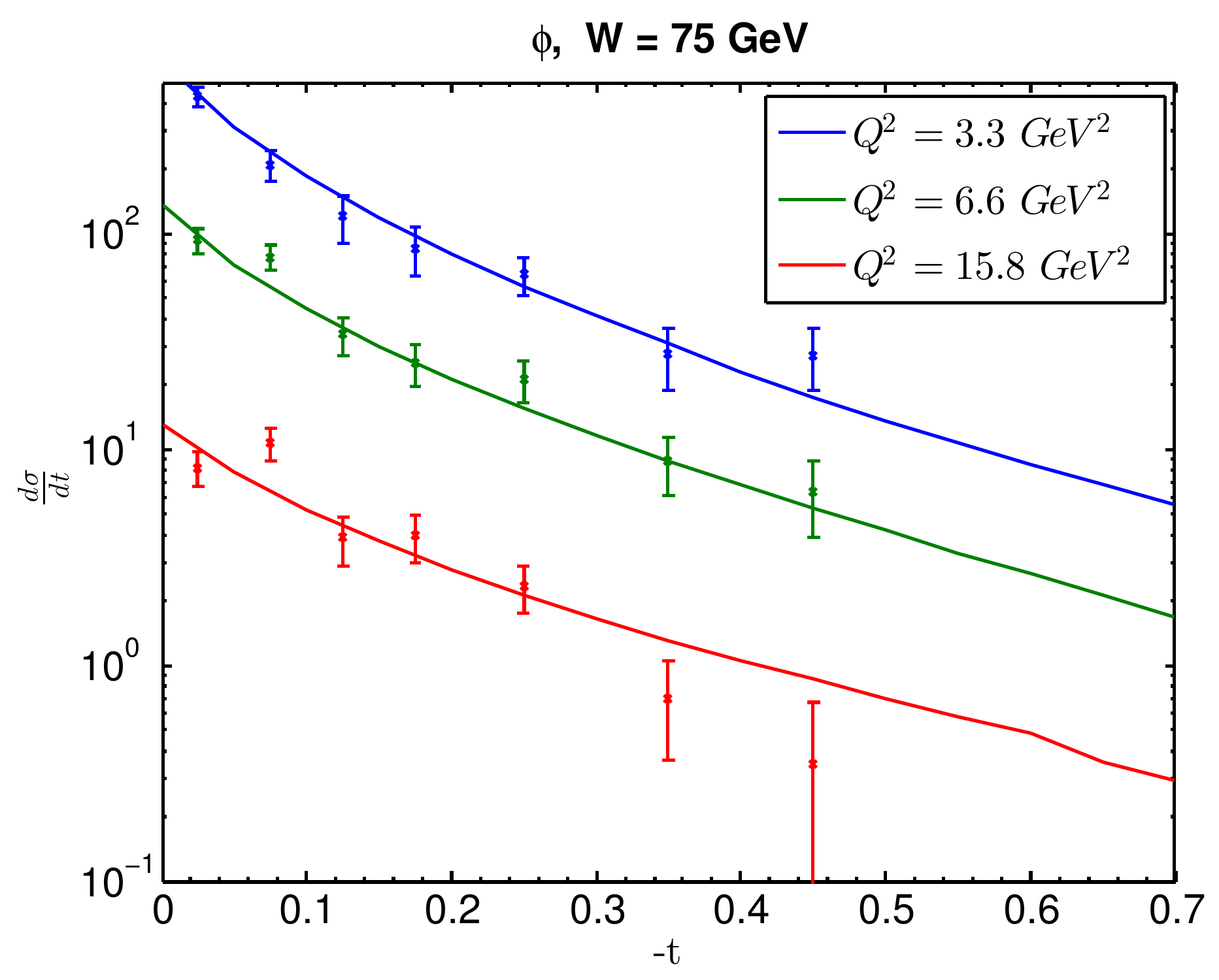}
\includegraphics[scale=0.425]{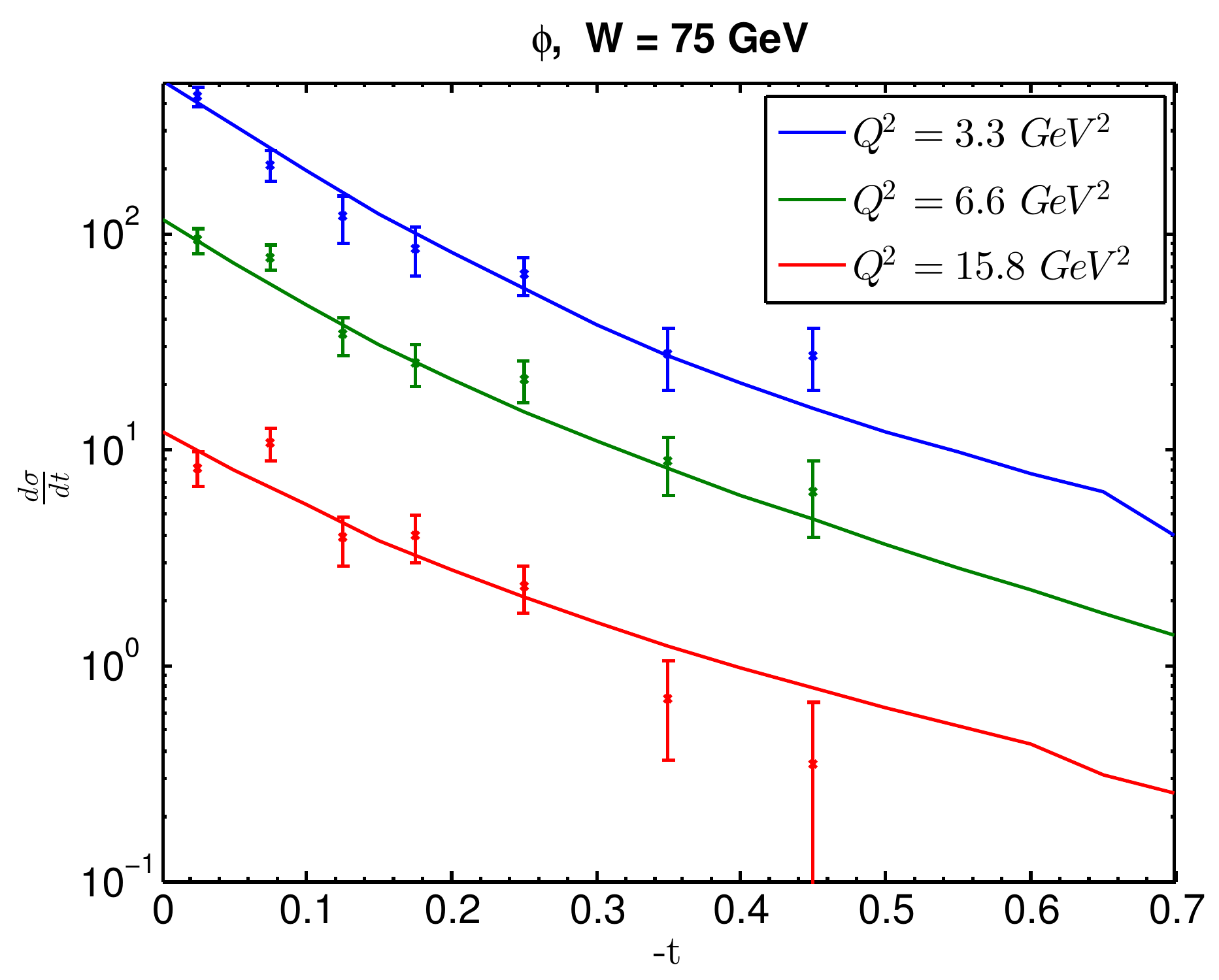}
\caption{{\small Differential cross section for the $\rho$ meson (top plots) and for the $\phi$ meson (bottom plots). 
Note that even though  the plots for each meson appear very similar, the left hand ones use the conformal Pomeron, and the right hand ones use the hard-wall Pomeron.}}
\label{fig:diff_rho_phi}
\end{center}
\end{figure}

Firstly, the fits to the full cross-sections provide very good $\chi^2 < 1$  in all cases. We display the fit to a representative sample of the data points\footnote{Of course, in Table 1 all of the values correspond to the fit to all the points.} in Figures \ref{fig:sig_conf} and \ref{fig:sig_hw}. Note the $\Omega$ production fit is only to 6 data points. The best fit for the intercept $j_0=2-\rho$ is in the range
$ 0.64 < \rho <0.76$ across the fits, which seems fairly stable,  
and consistent with the intercepts found in DIS \cite{Brower:2010wf} and DVCS \cite{Costa:2012fw}.
Note that, if we naively use the ${\cal N}=4$ SYM result $\rho=2/\sqrt{g^2_{YM}N}$,
this corresponds to fit values for $\alpha_s$ between 0.18 and 0.26. 
The best fit values for $g_0^2$ vary by almost an order of magnitude but since these are couplings specific to the meson states, that differ in each process,  there is therefore no expectation that they should precisely match. The best fit values for the position of the hard-wall in the proton wave function, $z_*$, vary by a factor of 4 or so across the fits (giving a typical value of  $500\ {\rm MeV}$ for the wall position).   The inclusion of the hard-wall cut-off parameter $z_0$ has very little impact on the goodness of fit in any of the cases (a fairly stable value around $200\  {\rm MeV}$ emerges).

\begin{figure}[t!]
\begin{center}
\includegraphics[scale=0.425]{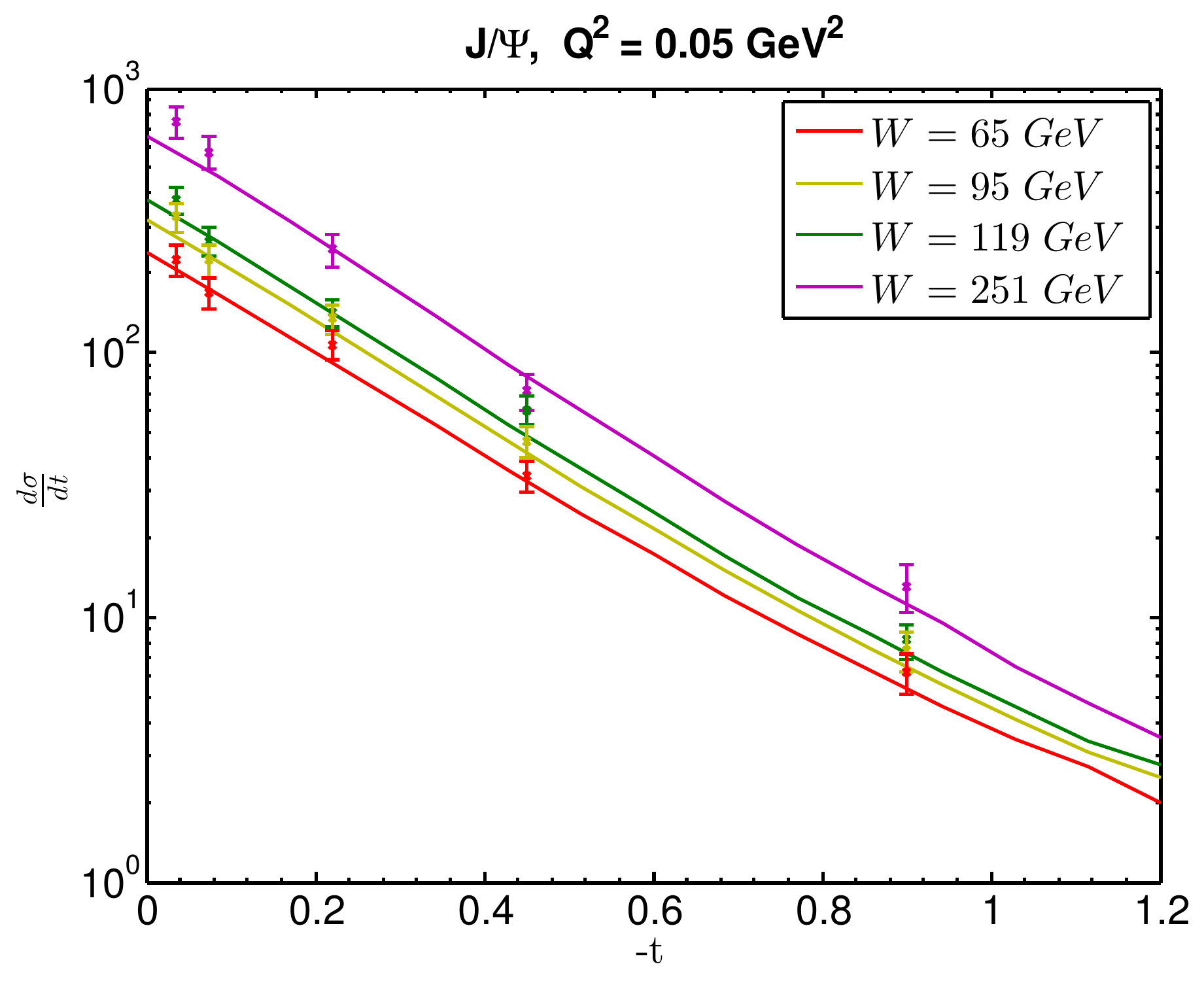}
\includegraphics[scale=0.425]{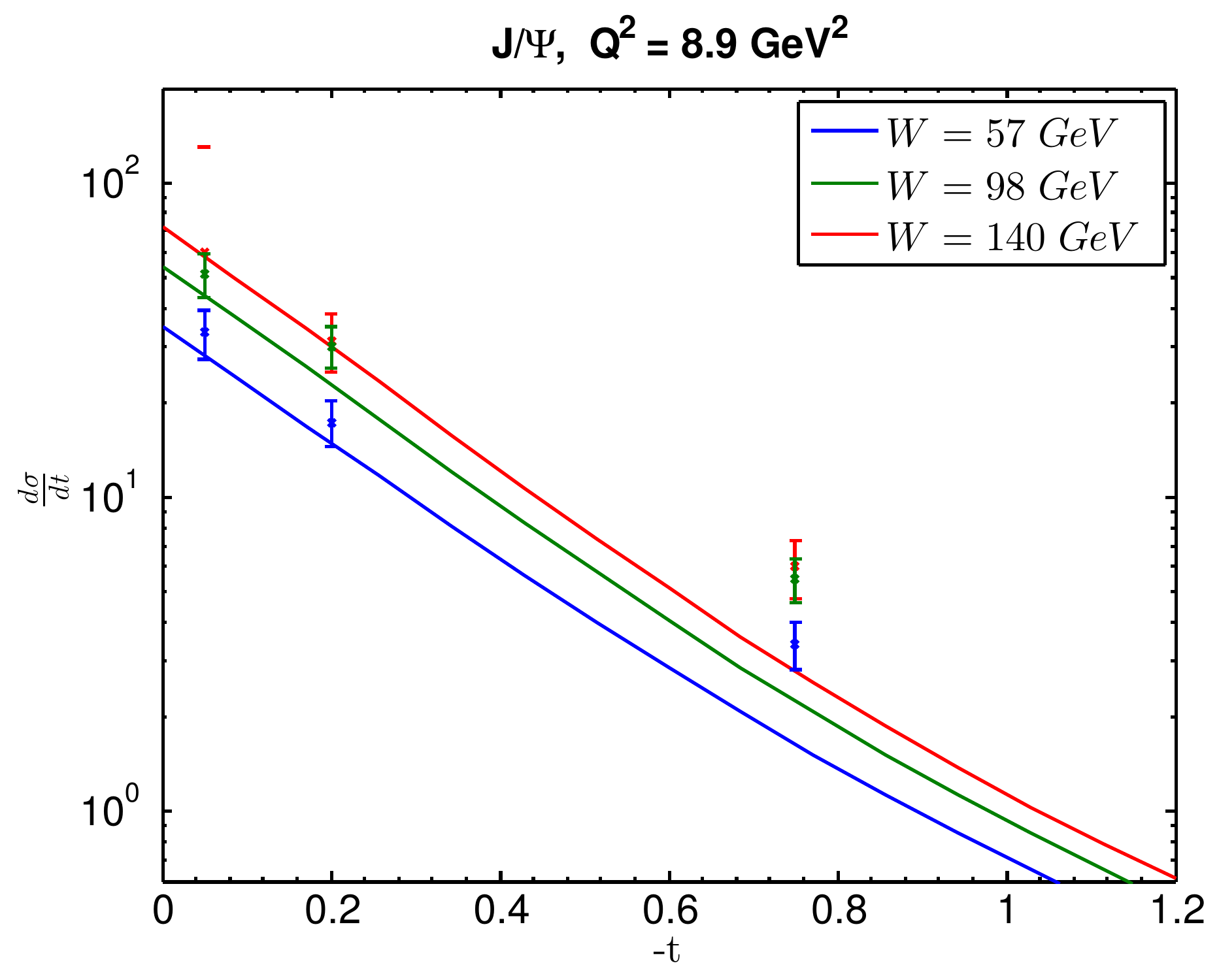}\\
\includegraphics[scale=0.425]{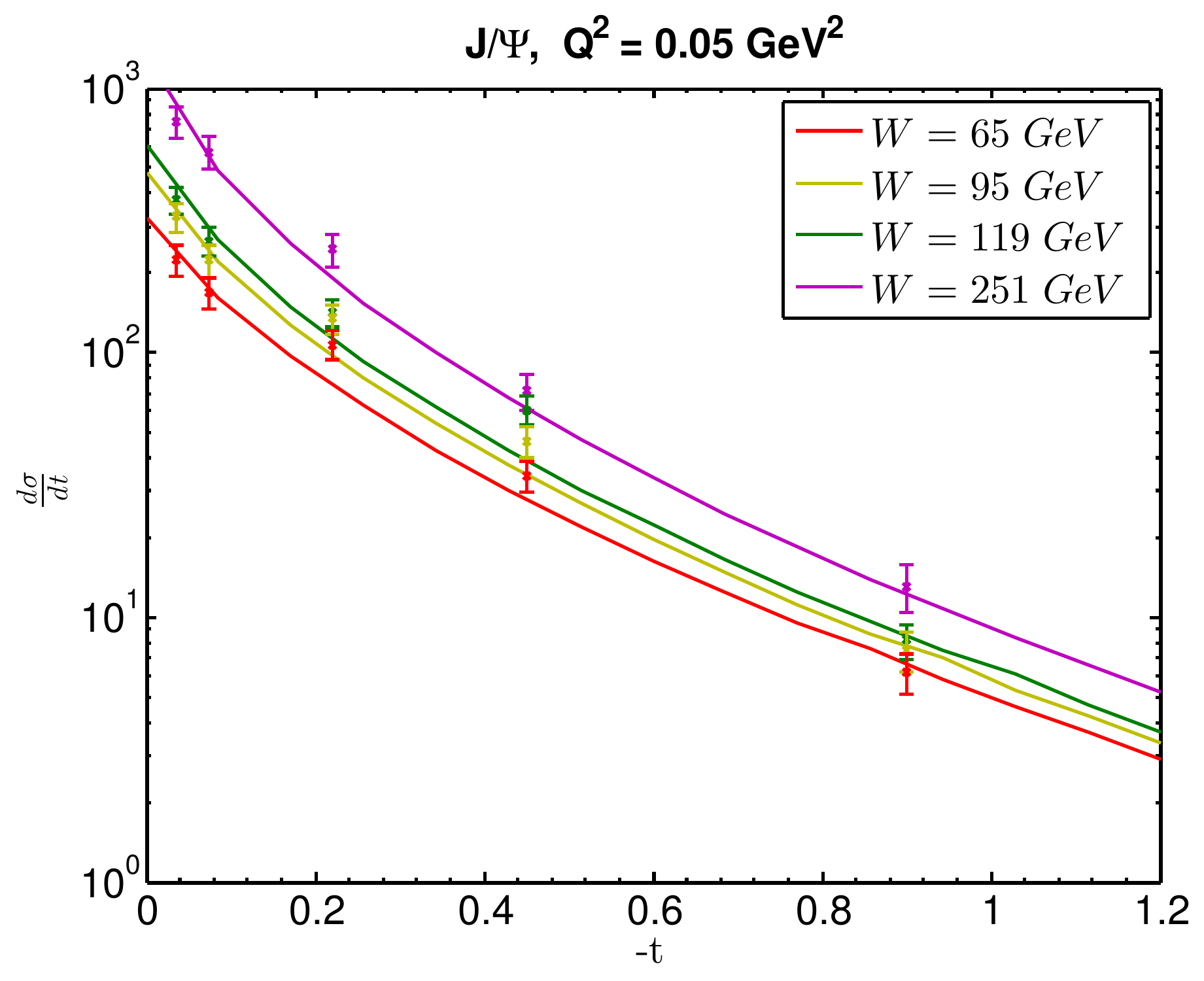}
\includegraphics[scale=0.425]{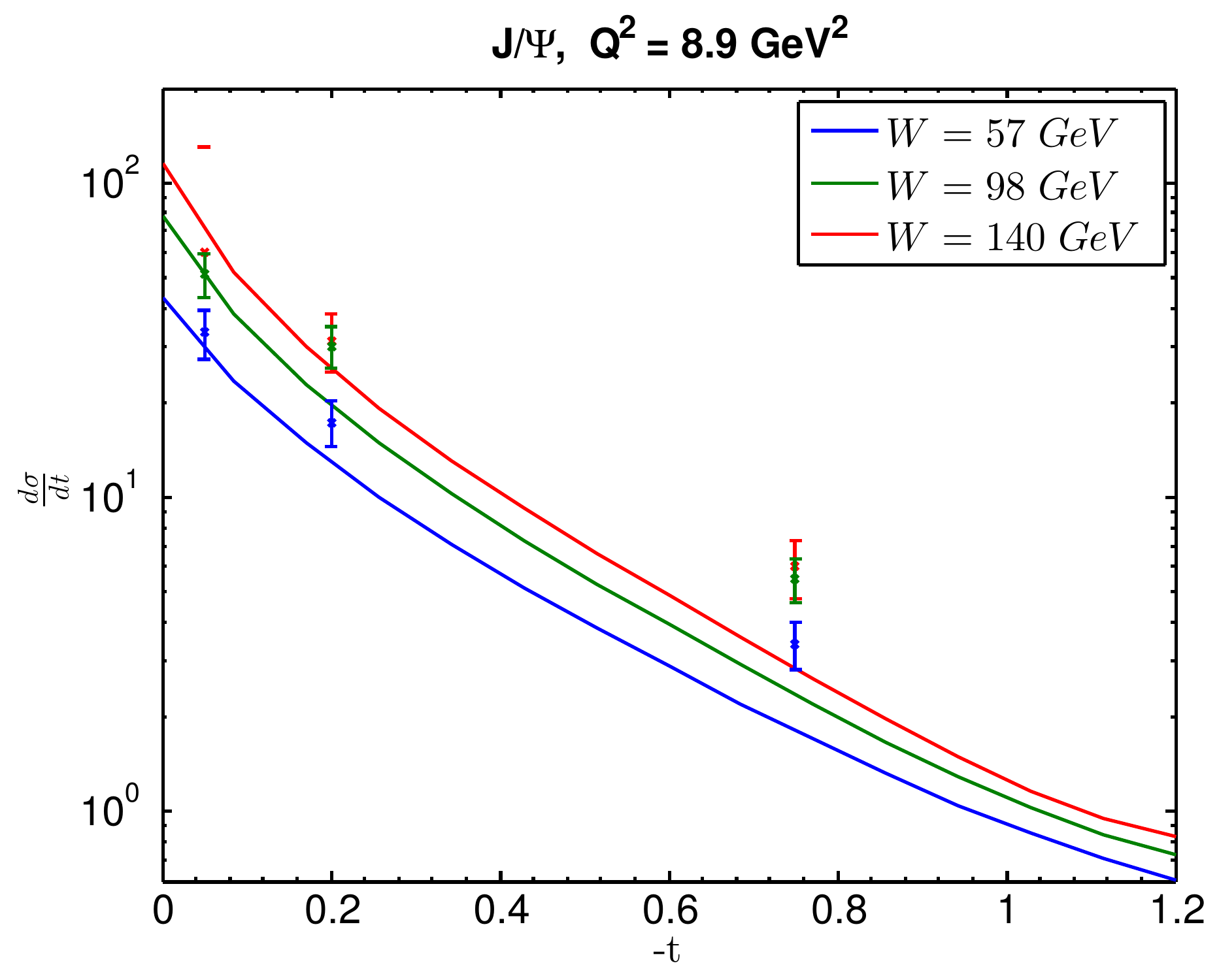}
\caption{{\small Differential cross section for the $J/\Psi$ meson for some of the $W$ and $Q^2$ values, using the hard-wall model (top) and the conformal model (bottom).}}
\label{fig:diff_jpsi}
\end{center}
\end{figure}

We can now turn to our fits of differential cross-section data (which is not available for the $\Omega$ meson).
To avoid cluttering the paper with too many figures, we display the fit to a representative sample of the data points in Figures \ref{fig:diff_rho_phi} and \ref{fig:diff_jpsi}. 
The fits are less good than
for the full cross-section data but still have $\chi^2 < 2$ in each case. To claim such a good fit for the $J/\Psi$ meson we do need to include the hard-wall parameter $z_0$ and this is the only place in our fits where it makes a significant impact.  For this process the momentum transfer energies $t$ go as low as $0.05\ {\rm GeV}^2$, which is already below the hard-wall cut off scale set by $1/z_0$. We therefore might need to improve the hard-wall model in order to obtain a better fit for this meson.  We also note that these fits are not quite as good as the equivalent ones to DIS and DVCS data using the AdS methods, presumably reflecting the additional complication of fitting the mesonic wave functions holographically. The fit parameters show the same broad behaviour as for the full cross-section although $z^*$ seems more stable.

In conclusion we find that the strong coupling AdS/CFT inspired model of low $x$  vector meson production gives 
a very good fit to the data, providing further evidence for the strength of gauge gravity duality methods. 


\bigskip

\begin{center} 
{\bf Acknowledgements} 
\end{center}
The authors are grateful to  Jo\~ao Penedones and Chung-I Tan for helpful discussions. This work was partially funded by grants PTDC/FIS/099293/2008 and CERN/FP/116358/2010. 
\emph{Centro de F\'{i}sica do Porto} is partially funded by FCT. The work of M.D. is supported by the FCT/Marie Curie Welcome II program. N.E. is grateful for the support of an STFC rolling grant.
The research leading to these results has received funding from the People Programme (Marie Curie Actions) of the European Union's Seventh Framework Programme FP7/2007-2013/ under REA Grant Agreements No 269217 and No 317089.

\bibliographystyle{./utphys}
\bibliography{./mybib}

\end{document}